\documentclass[12pt,english,aps,manuscript,12pt,nofootinbib]{revtex4}
\usepackage[T1]{fontenc}
\usepackage[latin9]{inputenc}
\usepackage{color}
\usepackage{babel}
\usepackage{amstext}
\usepackage{amssymb}
\usepackage{graphicx}
\usepackage{esint}
\usepackage[unicode=true,pdfusetitle,
 bookmarks=true,bookmarksnumbered=false,bookmarksopen=false,
 breaklinks=false,pdfborder={0 0 1},backref=false,colorlinks=true]
 {hyperref}
\hypersetup{
 pdfborderstyle=,linkcolor=blue,citecolor=cyan}

\makeatletter
\@ifundefined{textcolor}{}
{%
 \definecolor{BLACK}{gray}{0}
 \definecolor{WHITE}{gray}{1}
 \definecolor{RED}{rgb}{1,0,0}
 \definecolor{GREEN}{rgb}{0,1,0}
 \definecolor{BLUE}{rgb}{0,0,1}
 \definecolor{CYAN}{cmyk}{1,0,0,0}
 \definecolor{MAGENTA}{cmyk}{0,1,0,0}
 \definecolor{YELLOW}{cmyk}{0,0,1,0}
}

\usepackage{babel}

\usepackage{xcolor}

\usepackage{slashed}
\usepackage{braket}

\makeatother

\begin{document}
\title{Reexamination of local spin polarization beyond global equilibrium
\\
 in relativistic heavy ion collisions }
\author{Cong Yi}
\email{congyi@mail.ustc.edu.cn}

\affiliation{Department of Modern Physics, University of Science and Technology
of China, Hefei, Anhui 230026, China}
\author{Shi Pu}
\email{shipu@ustc.edu.cn}

\affiliation{Department of Modern Physics, University of Science and Technology
of China, Hefei, Anhui 230026, China}
\author{Di-Lun Yang}
\email{dlyang@gate.sinica.edu.tw}

\affiliation{Institute of Physics, Academia Sinica, Taipei 11529, Taiwan}
\begin{abstract}
We study local spin polarization in the relativistic hydrodynamic
model. Generalizing the Wigner functions previously obtained from
chiral kinetic theory by Y. Hidaka \textit{et al}. {[}Phys. Rev. D
97, 016004 (2018){]} to the massive case, we present the possible
contributions up to the order of $\hbar$ from thermal vorticity,
shear viscous tensor, other terms associated with the temperature
and chemical-potential gradients, and electromagnetic fields to the
local spin polarization. We then implement the (3+1)-dimensional viscous
hydrodynamic model to study the spin polarizations from these sources
with a small chemical potential and ignorance of electromagnetic fields
by adopting an equation of state different from those in other recent
studies. Although the shear correction alone upon the local polarization
results in a sign and azimuthal-angle dependence more consistent with
experimental observations, as also discovered in other recent studies,
it is mostly suppressed by the contributions from thermal vorticity
and other terms that yield an opposite trend. It is found that the
total local spin polarization can be very sensitive to the equation
of states, the ratio of shear viscosity to entropy density, and the
freeze-out temperature. 
\end{abstract}
\maketitle

\section{Introduction}

In noncentral heavy ion collisions, large orbital angular momenta
can be generated and transferred to the quark gluon plasma (QGP) in
the form of vorticity fields. Proposed by Liang and Wang \citep{ZTL_XNW_2005PRL,ZTL_XNW_2005PLB}
in early pioneer works, the global orbital angular momenta could trigger
the spin polarization of interacting partons through spin-orbit coupling.
The magnitude of vorticity can be accordingly extracted through the
spin polarization of hadrons created from the experiments. Later,
relativistic fermions with spin at thermal equilibrium were systematically
studied in the statistical model and a more precise relation to the
possible experimental observation was established \citep{Becattini:2007nd,Becattini:2007sr,Becattini:2013fla}.
On the other hand, the properties and dynamical evolution of vorticity
are further analyzed via numerical simulations in AMPT and HIJING
\citep{Jiang:2016woz,Deng_2016PRC,LiHui_prc2017_lpwx}. Assuming the
global equilibrium condition and employing the modified Cooper-Frye
formula for spin polarization \cite{Becattini:2013fla,Fang_2016PRC_polar},
the spin polarization of $\Lambda$ hyperons in heavy ion collisions
has been estimated based on UrQMD \citep{WeiDexian_prc2019_wdh} and
hydrodynamic models \citep{Csernai:2013bqa,Becattini:2013vja,Becattini:2015ska,Pang:2016igs,Wu:2020yiz}.
See also Refs.~\citep{Betz:2007kg,Ivanov_prc2017_is,Ivanov_prc2018_is,Ivanov_prc2019_its}
for other related studies. The theoretic predictions and follow-up
studies \citep{Karpenko_epjc2017_kb,XieYilong_prc2017_xwc,LiHui_prc2017_lpwx,Sun:2017xhx,WeiDexian_prc2019_wdh,Shi_plb2019_sll}
remarkably agree with the later measurement of the global polarization
for $\Lambda$ and $\bar{\Lambda}$ hyperons by the STAR Collaboration
\citep{STAR:2017ckg}. It is also shown that the average angular velocity
or vorticity of QGP is as large as $\omega\sim10^{22}s^{-1}$ \citep{STAR:2017ckg},
which reveals that the QGP could be the most vortical fluid so far.

Despite the agreement between theory and experiment on global polarization,
the STAR Collaboration has also measured the azimuthal angle dependence
of the spin polarization along the beam and the out-of-plane directions
in Au+Au collisions at $200$ GeV \citep{Adam:2019srw,Niida:2018hfw},
known as the longitudinal and transverse local spin polarization,
respectively. 
Most theoretical simulations, e.g., relativistic hydrodynamics \citep{Becattini_prl2018_bk,Fu:2020oxj}
and transport models \citep{XieYilong_prc2017_xwc,Xia:2018tes,WeiDexian_prc2019_wdh},
have failed to describe the measurements of local polarization, with
a few exceptions, such as the numerical simulation from the kinetic
theory of massless fermions in Refs.~\citep{Liu:2019krs} and results
from some phenomenological models in Ref.~\citep{Voloshin_2018epjWeb, Wu:2019eyi,Wu:2020yiz}.
In general, these theoretical estimations have found local polarization
with an opposite sign versus the experimental result. Also, feed-down
effects are found to be negligible for recoiling the tension \citep{Xia:2019fjf,Becattini_epjc2019_bcs}.
This disagreement between theory and experiment for local spin polarization
is dubbed the ``sign'' problem. 

Since most of the previous simulations have relied on the assumption
that spin polarization for $\Lambda$ hyperons at the freeze-out hypersurface
is mainly induced by the thermal vorticity in light of the statistical
model \citep{Becattini:2013fla} and Wigner-function approach \citep{Fang:2016uds}
at global equilibrium, it is generally believed that non-equilibrium
effects beyond the assumption of global equilibrium are essential
to delineate the local spin polarization. Consequently, many theoretical
efforts have focused on studying dynamical spin polarization. One
microscopic theory for tracking the dynamical evolution of spin transport
for relativistic fermions is the quantum kinetic theory (QKT). The
QKT was developed for massless fermions \citep{Hidaka:2017auj,Stephanov:2012ki,Son:2012zy,Chen:2012ca,Manuel:2013zaa,Manuel:2014dza,Chen:2014cla,Chen:2015gta,Hidaka:2016yjf,Mueller:2017lzw,Hidaka:2018mel,Huang:2018wdl,Gao:2018wmr,Liu:2018xip,Lin:2019ytz,Lin:2019fqo,Yamamoto:2020zrs}
at the very beginning, then known as the chiral kinetic theory (CKT).
To describe the spin dynamics, in particular, of strange (\textit{s})
quarks or $\Lambda$ hyperons, QKT has been extended to the case of
massive fermions \citep{Gao:2019znl,Weickgenannt:2019dks,Hattori:2019ahi,Wang:2019moi,Yang:2020hri,Weickgenannt:2020aaf,Weickgenannt:2020sit,Li:2019qkf,Liu:2020flb,Wang:2020pej,Weickgenannt:2021cuo,Sheng:2021kfc,Wang:2021qnt}.
In Ref. \citep{Zhang:2019xya}, another microscopic model for spin
polarization through particle collisions is proposed. As for macroscopic
descriptions, one may need to add the spin effects to the relativistic
hydrodynamics, i.e., relativistic spin hydrodynamics \citep{Florkowski_prc2018_ffjs,Florkowski:2018myy,Yang:2018lew,Becattini_plb2019_bfs,Florkowski:2018fap,Hattori:2019lfp,Bhadury:2020puc,Shi:2020htn,Montenegro:2017lvf,Montenegro:2017rbu,Li:2020eon}.
Spin hydrodynamic has been derived from many approaches, e.g. the
entropy principle \citep{Hattori:2019lfp,Fukushima:2020qta,Fukushima:2020ucl},
the Lagrangian effective theory \citep{Montenegro:2017lvf,Montenegro:2017rbu},
kinetic approaches \citep{Florkowski_prc2018_ffjs,Florkowski:2018myy,Becattini_plb2019_bfs,Florkowski:2018fap,Bhadury:2020puc,Shi:2020htn},
and general discussion from field theory \citep{Gallegos:2021bzp}.
Also, see recent reviews \citep{Wang:2017jpl,Becattini:2020ngo,Becattini:2020sww,Gao:2020vbh,Liu:2020ymh}
and the references therein.

Nevertheless, simulating the spin polarization of relativistic fermions
far from equilibrium is technically difficult and computationally
expensive. It is instead natural to explore the spin polarization
near local equilibrium. 
From the detailed balance of massless fermions with two-to-two scattering
in chiral kinetic theory, it has been shown that the Wigner function
pertinent to spin polarization contains several corrections besides
the thermal vorticity at local equilibrium as derived by some of the
authors of this paper \cite{Hidaka:2017auj} in 2017. Recently, the
authors of Refs. \citep{Liu:2020dxg,Liu:2021uhn} have found similar
contributions for massive fermions at local equilibrium, which have
a smooth connection to part of the massless result. In Ref. \cite{Becattini:2021suc},
the authors have simultaneously derived the shear-induced polarization
through statistical models. 
Based on their findings with particularly the shear correction, the
hydrodynamic simulations in Ref. \citep{Fu:2021pok} show qualitative
agreement with experimental data obtained by computing the polarization
of $s$ quarks in $\Lambda$ hyperons. Simultaneously, the authors
of Ref. \citep{Becattini:2021iol} have added the contribution from
the shear viscous tensor to the local spin polarization near isothermal
equilibrium, and the numerical results also agree with the experimental
data qualitatively.

A natural question then arises: Is the numerical finding above with
the shear corrections sensitive to the parameters chosen in the hydrodynamics
simulations? If the answer is positive, it implies that the correct
''sign'' for local polarization may not be solely attributed to
the shear correction at local equilibrium and further non-equilibrium
corrections depending on interaction should be considered. Therefore,
we follow the early work \citep{Hidaka:2017auj} done by some of the
authors of the present paper and list all possible local-equilibrium
corrections to the local spin polarization up to order $\hbar$. We
then implement $(3+1)$-dimensional viscous hydrodynamics to investigate
the polarization induced by these effects. In order to examine the
dependence of the numerical parameters in simulations, we use an equation
of state (EoS) different from the one adopted in Ref. \citep{Fu:2021pok}
and discuss the dependence of the freeze-out temperature and the ratio
of shear viscosity to entropy density.

The structure of the paper is as follows. In Sec. \ref{sec:Theoretical-analysis-from},
we review the main results of our early work \citep{Hidaka:2017auj}
and present all possible corrections at local equilibrium to polarization
up to order $\hbar$ explicitly. Next, we implement viscous hydrodynamical
simulations with the AMPT initial condition to study the polarization
and discuss the dependence of the freeze-out temperature and the ratio
of shear viscosity to entropy density in Sec. \ref{sec:Results-and-discussion}.
We summarize in Sec. \ref{sec:Conclusion}.

Through out this paper, we use the Minkowski metric $g_{\mu\nu}=\text{diag}\{+,-,-,-\}$
and define the Levi-Civita tensor $\epsilon^{\mu\nu\alpha\beta}$
with the convention $\epsilon^{0123}=-\epsilon_{0123}=1$. We also
introduce the notations $A_{(\rho}B_{\sigma)}\equiv(A_{\rho}B_{\sigma}+A_{\sigma}B_{\rho})/2$
and $A_{[\rho}B_{\sigma]}\equiv(A_{\rho}B_{\sigma}-A_{\sigma}B_{\rho})/2$.

\section{Theoretical analysis from quantum kinetic theories \label{sec:Theoretical-analysis-from}}

We are interested in the polarization (pseudo) vector characterized
by the axial-charge current density in phase space, 
\begin{equation}
\mathcal{J}_{5}^{\mu}(p,X)\equiv2\int_{p\cdot n}[\mathcal{J}_{+}^{\mu}(p,X)-\mathcal{J}_{-}^{\mu}(p,X)],\label{J5_density_def}
\end{equation}
where $\int_{p\cdot n}\equiv\int dp\cdot np\cdot n\theta(p\cdot n)/(2\pi)$
with $\theta(p\cdot n)$ a unit-step function \footnote{For computation of the spin polarization, we usually apply the on-shell
Wigner functions. Here we thus integrate over the energy defined as
$p\cdot n$, where $n^{\mu}$ may be chosen as the fluid four velocity
at thermal equilibrium. Also, we further introduce a unit-step function
to omit the contribution from anti-fermions.}. Here $\mathcal{J}_{+}^{\mu}(p,X)$ and $\mathcal{J}_{-}^{\mu}(p,X)$
denote the Wigner functions for right- and left-handed fermions, respectively.
Given $\mathcal{J}_{5}^{\mu}(p,X)$, we can calculate the spin polarization
of $\Lambda$ hyperons via the modified Cooper-Frye formula \cite{Becattini:2013fla},
\begin{equation}
\mathcal{S}^{\mu}({\bf p})=\frac{\int d\Sigma\cdot p\mathcal{J}_{5}^{\mu}(p,X)}{2m_{\Lambda}\int d\Sigma\cdot\mathcal{N}(p,X)},\label{eq:S_01}
\end{equation}
where $\mathcal{N}^{\mu}(p,X)\equiv2\int_{p\cdot n}[\mathcal{J}_{+}^{\mu}(p,X)+\mathcal{J}_{-}^{\mu}(p,X)]$
is the number density in phase space, $m_{\Lambda}$ is the mass of
$\Lambda$ and $\Sigma_{\mu}$ is the normal vector of the freeze-out
surface.

As derived in the early work \citep{Hidaka:2017auj}, the Wigner functions
for right- or left-handed fermions near local equilibrium are given
by \footnote{In general, there exist off-equilibrium corrections pertinent to collisions,
which are, however, higher order in gradient expansion starting from
$\mathcal{O}(\partial)$ and $\mathcal{O}(\partial^{2})$ for even-
and odd-parity terms, respectively. These corrections are neglected
here.} 
\begin{eqnarray}
\mathcal{J}_{\lambda}^{\mu}(p,X) & = & 2\pi\textrm{sign}(u\cdot p)\left\{ \delta(p^{2})p^{\mu}+\lambda\frac{\hbar}{2}\delta(p^{2})\left[u^{\mu}(p\cdot\omega)-\omega^{\mu}(u\cdot p)\right.\right.\nonumber \\
 &  & \left.\left.-2S_{(u)}^{\mu\nu}\tilde{E}_{\nu}\right]\partial_{u\cdot p}+\lambda\frac{\hbar}{4}\epsilon^{\mu\nu\alpha\beta}F_{\alpha\beta}\partial_{\nu}^{p}\delta(p^{2})\right\} f_{\lambda}^{(0)},\label{eq:J_01}
\end{eqnarray}
where $\lambda=\pm$ for right- and left- handed fermions, $u^{\mu}$
is the fluid four velocity, 
\begin{eqnarray}
S_{(u)}^{\mu\nu} & = & \epsilon^{\mu\nu\alpha\beta}p_{\alpha}u_{\beta}/(2u\cdot p),\nonumber \\
\tilde{E}_{\nu} & = & E_{\nu}+T\partial_{\nu}\frac{\mu_{\lambda}}{T}+\frac{(u\cdot p)}{T}\partial_{\nu}T-p^{\sigma}[\partial_{<\sigma}u_{\nu>}+\frac{1}{3}\Delta_{\sigma\nu}(\partial\cdot u)+u_{\nu}Du_{\sigma}].
\end{eqnarray}
and 
\begin{equation}
f_{\lambda}^{(0)}=1/(e^{(u\cdot p-\mu_{\lambda})/T}+1),\label{equil_dist}
\end{equation}
is the distribution function with $T$ the local temperature, and
$\mu_{\pm}$ the chemical potentials for right- or left-handed fermions,
respectively. When evaluating the current by integrating $\mathcal{J}_{\lambda}^{\mu}(p,X)$
over $p$, we may rewrite the last term in Eq.~(\ref{eq:J_01}) as
\begin{eqnarray}
2\pi\textrm{sign}(u\cdot p)\lambda\frac{\hbar}{4}\epsilon^{\mu\nu\alpha\beta}F_{\alpha\beta}\partial_{\nu}^{p}\delta(p^{2})f_{\lambda}^{(0)}\approx-2\pi\textrm{sign}(u\cdot p)\lambda\frac{\hbar\delta(p^{2})}{4}\epsilon^{\mu\nu\alpha\beta}F_{\alpha\beta}\partial_{p\nu}f_{\lambda}^{(0)},
\end{eqnarray}
by dropping the surface term. The rewritten form allows us to properly
introduce the axial-charge current density in phase space through
Eq.~(\ref{J5_density_def}). In general, $f_{\lambda}$ should incorporate
non-equilibrium corrections depending on interaction to satisfy the
kinetic equation near local equilibrium, while the corrections are
expected to be higher order in the gradient expansion and are omitted
here for simplicity \cite{Hidaka:2017auj,Hidaka:2018ekt}. Here, electromagnetic
fields are defined in the fluid rest frame, 
\begin{equation}
E_{\mu}\equiv u^{\nu}F_{\mu\nu},\quad B^{\mu}\equiv\frac{1}{2}\epsilon^{\mu\nu\alpha\beta}u_{\nu}F_{\alpha\beta}.
\end{equation}
We also decompose the derivative of $u_{\nu}$ as 
\begin{eqnarray}
\partial_{\mu}u_{\nu} & = & \partial_{<\mu}u_{\nu>}+u_{\mu}Du_{\nu}+\frac{1}{3}\Delta_{\mu\nu}(\partial\cdot u)+\omega_{\mu\nu},\label{decompose_u_01}
\end{eqnarray}
where 
\begin{eqnarray}
\Delta^{\mu\nu} & = & g^{\mu\nu}-u^{\mu}u^{\nu},
\end{eqnarray}
is the projector, 
\begin{equation}
D\equiv u\cdot\partial,
\end{equation}
$A^{<\mu\nu>}$ is the traceless part of an arbitrary tensor $A^{\mu\nu}$,
\begin{equation}
A^{<\mu\nu>}\equiv\frac{1}{2}[\Delta^{\mu\alpha}\Delta^{\nu\beta}+\Delta^{\nu\beta}\Delta^{\mu\alpha}]A_{\alpha\beta}-\frac{1}{3}\Delta^{\mu\nu}\Delta^{\alpha\beta}A_{\alpha\beta},
\end{equation}
and $\omega^{\mu\nu}$ is the vorticity tensor 
\begin{eqnarray}
\omega_{\alpha\beta} & = & \epsilon_{\alpha\beta\mu\nu}u^{\mu}\omega^{\nu}+\frac{1}{2}(u_{\alpha}Du_{\beta}-u_{\beta}Du_{\alpha}),
\end{eqnarray}
with vorticity defined as, 
\begin{equation}
\omega^{\mu}=\frac{1}{2}\epsilon^{\mu\nu\alpha\beta}u_{\nu}\partial_{\alpha}u_{\beta}.
\end{equation}

Using the relation, 
\begin{eqnarray}
u^{\mu}(p\cdot\omega)-\omega^{\mu}(u\cdot p) & = & -\frac{1}{2}\epsilon^{\mu\nu\alpha\beta}p_{\nu}\partial_{\alpha}u_{\beta}+\frac{1}{2}\epsilon^{\mu\nu\alpha\beta}p_{\nu}u_{\alpha}Du_{\beta},
\end{eqnarray}
we find 
\begin{eqnarray}
 &  & u^{\mu}(p\cdot\omega)-\omega^{\mu}(u\cdot p)-2S_{(u)}^{\mu\nu}\tilde{E}_{\nu}\nonumber \\
 & = & -\frac{1}{2}\epsilon^{\mu\nu\alpha\beta}p_{\nu}T\partial_{\alpha}\frac{u_{\beta}}{T}+\frac{1}{2}\epsilon^{\mu\nu\alpha\beta}p_{\nu}u_{\alpha}Du_{\beta}-\frac{1}{2}\epsilon^{\mu\nu\alpha\beta}p_{\alpha}u_{\beta}\frac{1}{T}\partial_{\nu}T\nonumber \\
 &  & -\frac{1}{(u\cdot p)}\epsilon^{\mu\nu\alpha\beta}p_{\alpha}u_{\beta}E_{\nu}-\frac{T}{(u\cdot p)}\epsilon^{\mu\nu\alpha\beta}p_{\alpha}u_{\beta}\partial_{\nu}\frac{\mu_{\lambda}}{T}+\frac{1}{(u\cdot p)}\epsilon^{\mu\nu\alpha\beta}p_{\alpha}u_{\beta}p^{\sigma}\partial_{<\sigma}u_{\nu>}.
\end{eqnarray}

For convenience, we decompose $\mathcal{J}_{5}^{\mu}$ as, 
\begin{equation}
\mathcal{J}_{5}^{\mu}=\mathcal{J}_{\textrm{thermal}}^{\mu}+\mathcal{J}_{\textrm{shear}}^{\mu}+\mathcal{J}_{\textrm{accT}}^{\mu}+\mathcal{J}_{\textrm{chemical}}^{\mu}+\mathcal{J}_{\textrm{EB}}^{\mu},\label{J5_decomp}
\end{equation}
where 
\begin{eqnarray}
\mathcal{J}_{\textrm{thermal}}^{\mu} & = & a\frac{1}{2}\epsilon^{\mu\nu\alpha\beta}p_{\nu}\partial_{\alpha}\frac{u_{\beta}}{T},\nonumber \\
\mathcal{J}_{\textrm{shear}}^{\mu} & = & -a\frac{1}{(u\cdot p)T}\epsilon^{\mu\nu\alpha\beta}p_{\alpha}u_{\beta}p^{\sigma}\partial_{<\sigma}u_{\nu>}\nonumber \\
\mathcal{J}_{\textrm{accT}}^{\mu} & = & -a\frac{1}{2T}\epsilon^{\mu\nu\alpha\beta}p_{\nu}u_{\alpha}(Du_{\beta}-\frac{1}{T}\partial_{\beta}T).\nonumber \\
\mathcal{J}_{\textrm{chemical}}^{\mu} & = & a\frac{1}{(u\cdot p)}\epsilon^{\mu\nu\alpha\beta}p_{\alpha}u_{\beta}\partial_{\nu}\frac{\mu}{T},\nonumber \\
\mathcal{J}_{\textrm{EB}}^{\mu} & = & a\frac{1}{(u\cdot p)T}\epsilon^{\mu\nu\alpha\beta}p_{\alpha}u_{\beta}E_{\nu}+a\frac{B^{\mu}}{T},
\end{eqnarray}
with 
\begin{equation}
a=4\pi\hbar\textrm{sign}(u\cdot p)\delta(p^{2})f_{V}^{(0)}(1-f_{V}^{(0)}).\label{eq:A_01}
\end{equation}
Note that 
\begin{equation}
f_{V}^{(0)}=\frac{1}{2}(f_{+}^{(0)}+f_{-}^{(0)}),
\end{equation}
and we have set the same chemical potential for both left and right
fermions, $\mu=\mu_{+}=\mu_{-}$ for simplicity. The subscripts, ``thermal'',
``shear'', ``accT'', ``chemical'' and ``EB'', stand for the
terms related to the thermal vorticity, shear viscous tensor, fluid
acceleration minus the gradient of temperature $Du_{\mu}-(\Delta_{\mu\nu}\partial^{\nu}T)/T$,
gradient of $\mu/T$, and electromagnetic fields, respectively. Except
for $\mathcal{J}_{\textrm{thermal}}^{\mu}$ and part of $\mathcal{J}_{\textrm{EB}}^{\mu}$
led by magnetic fields, all other terms in $\mathcal{J}_{5}^{\mu}$
come from the corrections beyond global equilibrium.

Let us take a close look at $\mathcal{J}_{\textrm{accT}}^{\mu}$,
which is usually neglected for the following reason. One may utilize
hydrodynamic equations of motion up to the order of $\hbar$ or $\partial^{2}$,
\begin{eqnarray}
DT & = & -\frac{\epsilon+P}{p}\Delta^{\mu\nu}\partial_{\nu}u_{\mu}+\mathcal{O}(\hbar,\partial^{2}),\nonumber \\
Du_{\mu} & = & \frac{\Delta_{\mu\nu}\partial^{\nu}P}{\epsilon+P}+\mathcal{O}(\hbar,\partial^{2}),\nonumber \\
D\bar{\mu}_{{\rm R/L}} & = & \mathcal{O}(\hbar,\partial^{2}),\label{hydro_EOM}
\end{eqnarray}
to replace the temporal derivatives $D$ in thermodynamic parameters,
where $\epsilon$ and $P$ correspond to the energy density and pressure,
respectively. Here the $\hbar$ corrections in Eq.~(\ref{hydro_EOM})
are irrelevant since they only contribute to higher-order terms at
$\mathcal{O}(\hbar^{2})$ in Wigner functions except for the off-equilibrium
fluctuations led by collisions.

In the ideal limit, one of the hydrodynamic equations of motion becomes
$Du_{\mu}=(\Delta_{\mu\nu}\partial^{\nu}T)/T$. Therefore, $\mathcal{J}_{\textrm{accT}}^{\mu}$
vanishes in the ideal hydrodynamics. In realistic hydrodynamic simulations,
 dissipative corrections could further modify the evolution of $Du_{\mu}$.
In viscous hydrodynamics, we have, 
\begin{equation}
Du_{\alpha}=\frac{1}{\epsilon+P}\Delta_{\mu\alpha}\partial^{\mu}P-\frac{1}{\epsilon+P}\Delta_{\mu\alpha}\partial_{\nu}\pi^{\mu\nu}+\mathcal{O}(\hbar,\partial^{3})\approx\frac{\Delta_{\mu\alpha}}{T}\big(\partial^{\mu}T-s^{-1}\partial_{\nu}\pi^{\mu\nu}\big),
\end{equation}
where $\pi^{\mu\nu}=2\Delta^{\mu\alpha}\Delta^{\nu\beta}\eta(\partial_{(\alpha}u_{\beta)}-g_{\alpha\beta}\partial_{\rho}u^{\rho}/3)+\mathcal{O}(\partial^{2})$
corresponds to the shear-stress tensor with $\eta$ being the shear
viscosity and $s$ denoting the entropy density. Here we take $\epsilon+P=Ts$
and omit the correction from the bulk viscosity for simplicity. We
can further rewrite $\mathcal{J}_{\textrm{accT}}^{\mu}$ as 
\begin{eqnarray}
\mathcal{J}_{\textrm{accT}}^{\mu} & = & a\frac{1}{2T^{2}s}\epsilon^{\mu\nu\alpha\beta}p_{\nu}u_{\alpha}\partial^{\rho}\pi_{\beta\rho}\approx a\frac{\eta}{T^{2}s^{2}}\epsilon^{\mu\nu\alpha\beta}p_{\nu}u_{\alpha}\partial^{\rho}(s\hat{\pi}_{\beta\rho}),
\end{eqnarray}
where $\hat{\pi}^{\mu\nu}=\pi^{\mu\nu}/(2\eta)$ and we have assumed
that $\eta/s$ is a constant to make the final approximation above.
Consequently, we would like to emphasize here that the contributions
from $\mathcal{J}_{\textrm{accT}}^{\mu}$ to the local spin polarization
should depend on the EoS and parameter $\eta/s$. 
Note that in early work \citep{Karpenko:2018erl}, the authors discussed
the contribution from fluid acceleration $Du_{\beta}$ implicitly
involved in $\mathcal{J}_{\textrm{thermal}}^{\mu}$ to local polarization.
Additionally, the magnetic-field contribution in $\mathcal{J}_{\textrm{EB}}^{\mu}$
has also been discussed in Ref.~\cite{Fang_2016PRC_polar}.

Although we discuss the polarization of massless fermions above, we
can extend our analysis to massive fermions. In the case of massive
fermions, we can generalize the on-shell condition $\delta(p^{2})$
in Eq. (\ref{eq:A_01}) to $\delta(p^{2}-m_{i}^{2})$ with $m_{i}$
being the mass of fermions. Then, we obtain $\mathcal{S}^{\mu}$ in
Eq. (\ref{eq:S_01}) from different sources, 
\begin{eqnarray}
\mathcal{S}_{\textrm{thermal}}^{\mu}(\mathbf{p}) & = & \frac{\hbar}{8m_{\Lambda}N}\int d\Sigma^{\sigma}p_{\sigma}f_{V}^{(0)}(1-f_{V}^{(0)})\epsilon^{\mu\nu\alpha\beta}p_{\nu}\partial_{\alpha}\frac{u_{\beta}}{T},\nonumber \\
\mathcal{S}_{\textrm{shear}}^{\mu}(\mathbf{p}) & = & -\frac{\hbar}{4m_{\Lambda}N}\int d\Sigma\cdot pf_{V}^{(0)}(1-f_{V}^{(0)})\frac{\epsilon^{\mu\nu\alpha\beta}p_{\alpha}u_{\beta}}{(u\cdot p)T}\frac{1}{2}p^{\sigma}[(\partial_{\sigma}u_{\nu}+\partial_{\nu}u_{\sigma})-u_{\sigma}Du_{\nu}]\nonumber \\
\mathcal{S}_{\textrm{accT}}^{\mu}(\mathbf{p}) & = & -\frac{\hbar}{8m_{\Lambda}N}\int d\Sigma\cdot pf_{V}^{(0)}(1-f_{V}^{(0)})\frac{1}{T}\epsilon^{\mu\nu\alpha\beta}p_{\nu}u_{\alpha}(Du_{\beta}-\frac{1}{T}\partial_{\beta}T),\nonumber \\
\mathcal{S}_{\textrm{chemical}}^{\mu}(\mathbf{p}) & = & \frac{\hbar}{4m_{\Lambda}N}\int d\Sigma\cdot pf_{V}^{(0)}(1-f_{V}^{(0)})\frac{1}{(u\cdot p)}\epsilon^{\mu\nu\alpha\beta}p_{\alpha}u_{\beta}\partial_{\nu}\frac{\mu}{T},\nonumber \\
\mathcal{S}_{\textrm{EB}}^{\mu}(\mathbf{p}) & = & \frac{\hbar}{4m_{\Lambda}N}\int d\Sigma\cdot pf_{V}^{(0)}(1-f_{V}^{(0)})\left(\frac{1}{(u\cdot p)T}\epsilon^{\mu\nu\alpha\beta}p_{\alpha}u_{\beta}E_{\nu}+\frac{B^{\mu}}{T}\right),\label{eq:S_all}
\end{eqnarray}
where $N=\int d\Sigma^{\mu}p_{\mu}f_{V}^{(0)}$ and now $p^{0}=\sqrt{|\mathbf{p}|^{2}+m_{i}^{2}}$.
In this work, we only evaluate the spin polarization of $\Lambda$
and neglect $\bar{\Lambda}$.

We would like to emphasize that at global equilibrium it is necessary
to impose a vanishing shear viscous strength to satisfy the free-streaming
kinetic theory, which is equivalent to the Killing condition. On the
other hand, at local equilibrium, one can also decompose $\partial_{\mu}u_{\nu}$
into more terms as opposed to our decomposition in Eq. (\ref{decompose_u_01}),
while the total effect led by the fluid-velocity gradient at local
equilibrium should remain unchanged.

\section{Results and discussion \label{sec:Results-and-discussion}}


\subsection{Setup for Simulations}

In this section, we implement (3+1) dimensional viscous hydrodynamic
CLVisc \citep{Pang:2012he,Pang:2018zzo} with AMPT initial conditions
\citep{Lin:2004en,Wu:2019eyi,Wu:2020yiz} to generate the freeze-out
hyper-surface and the profile of the fluid velocity and temperature
at that hyper-surface. Unless noted otherwise, we choose the EoS ``\emph{s95p-pce}''
\citep{Huovinen:2009yb} instead of the EoS \citep{Denicol:2018wdp}
used in Refs. \citep{Fu:2020oxj,Fu:2021pok}.

\begin{figure}
\includegraphics[scale=0.4]{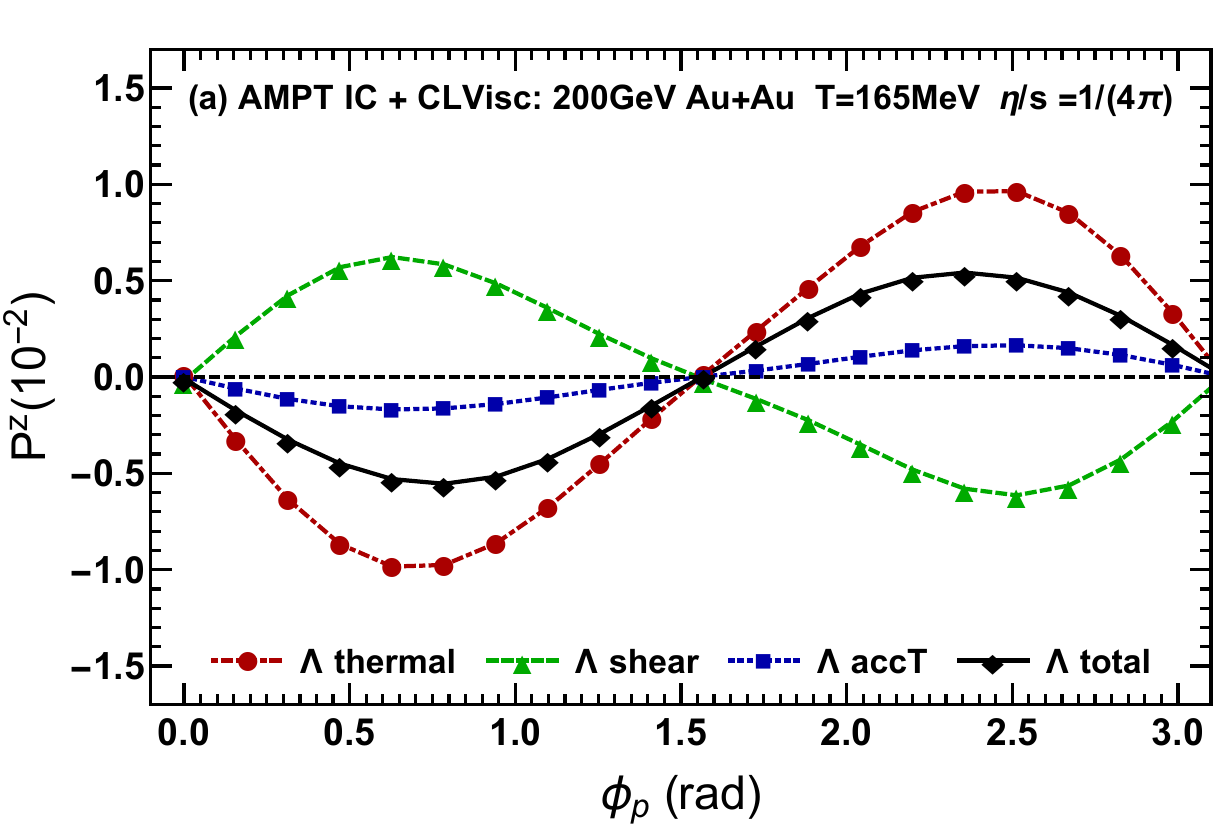}\includegraphics[scale=0.4]{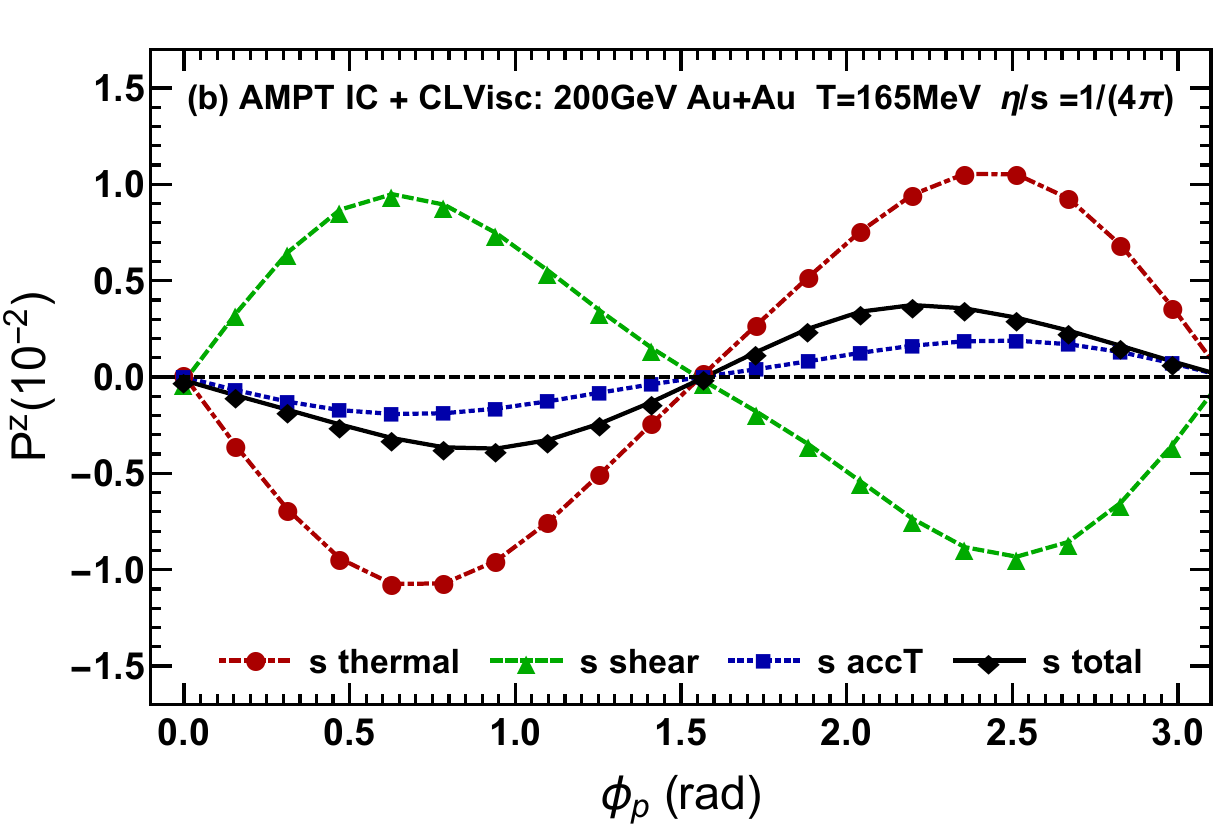}

\includegraphics[scale=0.4]{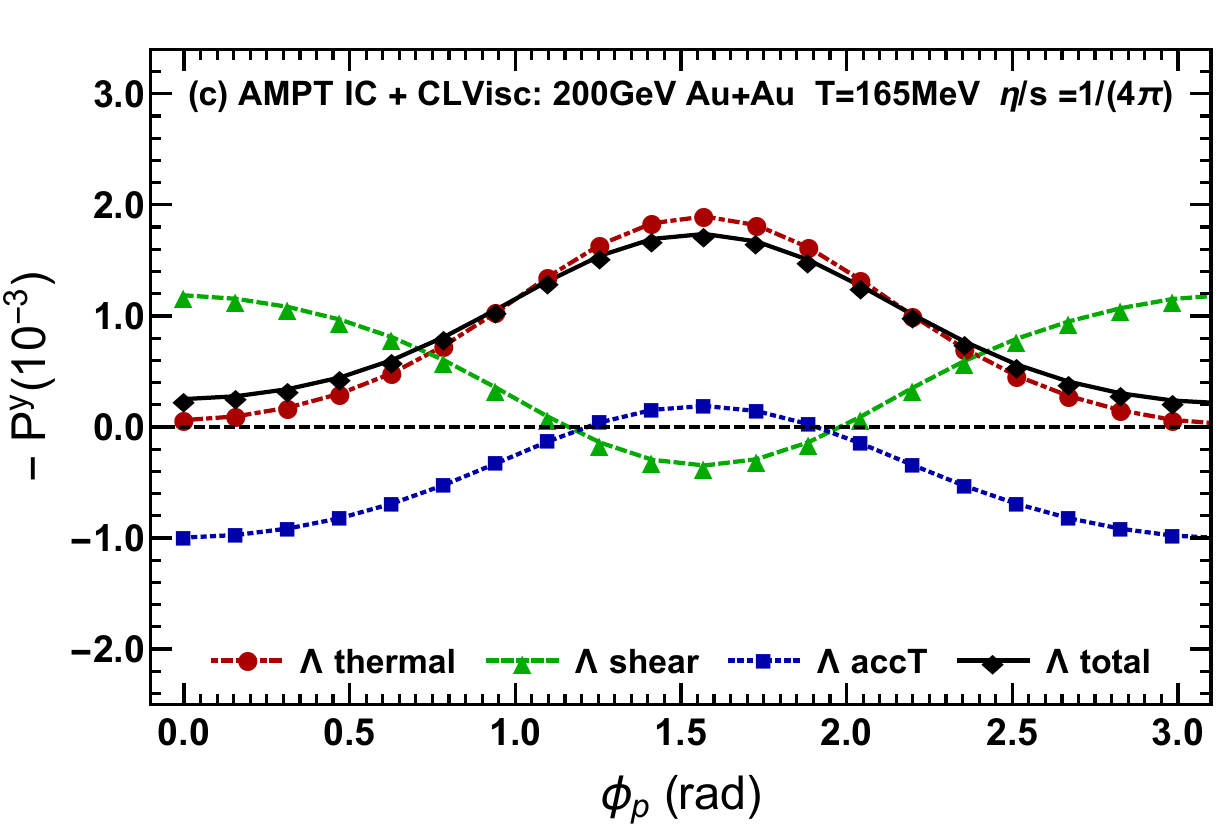}\includegraphics[scale=0.4]{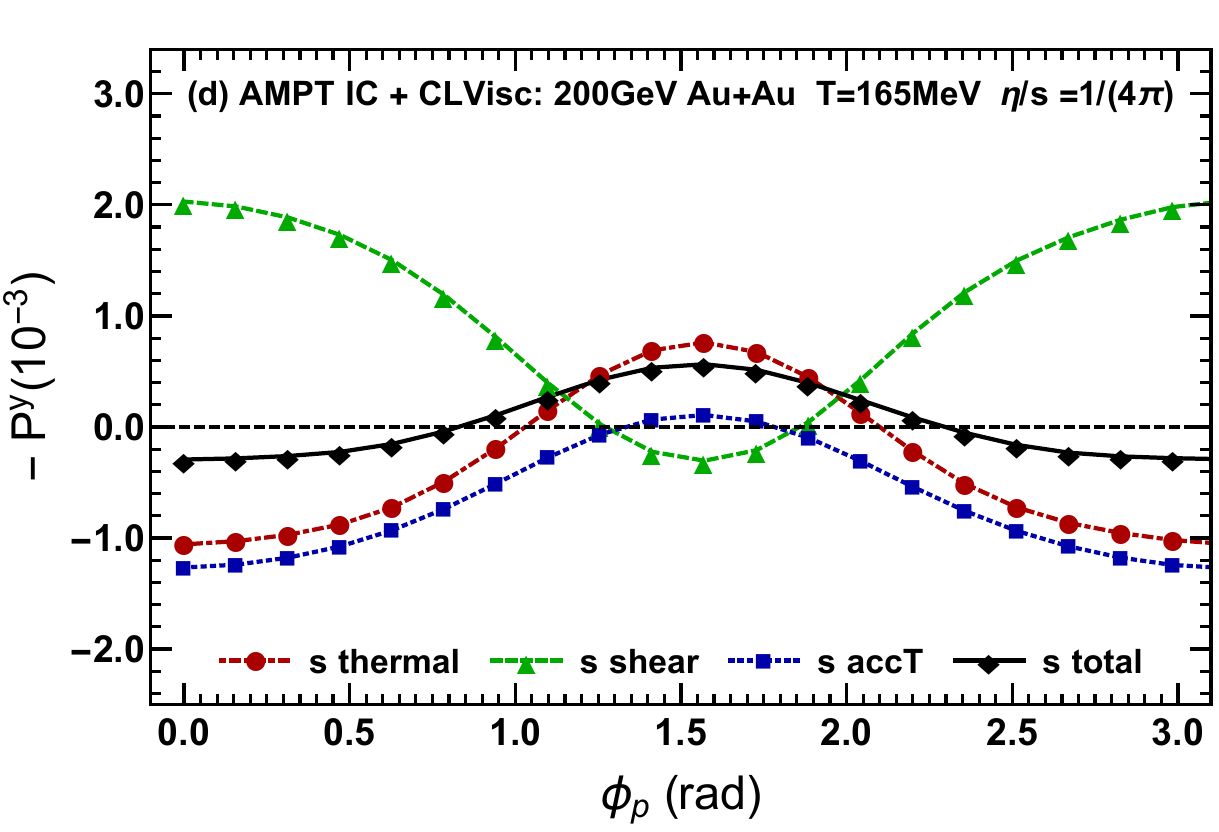}

\caption{The polarization $\mathcal{P}^{z}$ and $\mathcal{P}^{y}$ as a function
of $\phi_{p}$ for $\Lambda$ (a,c) and $s$ (b,d) equilibrium scenarios.
We have chosen $\eta/s=1/(4\pi)$ and the freeze-out temperature is
$165\,\textrm{MeV}$. Red, green, and blue curves represent the contributions
from the thermal vorticity, shear-induced polarization, and acceleration
terms, respectively. The black curve denotes the total polarization.
\label{fig:PzPy_01} 
}
\end{figure}

\begin{figure}
\includegraphics[scale=0.35]{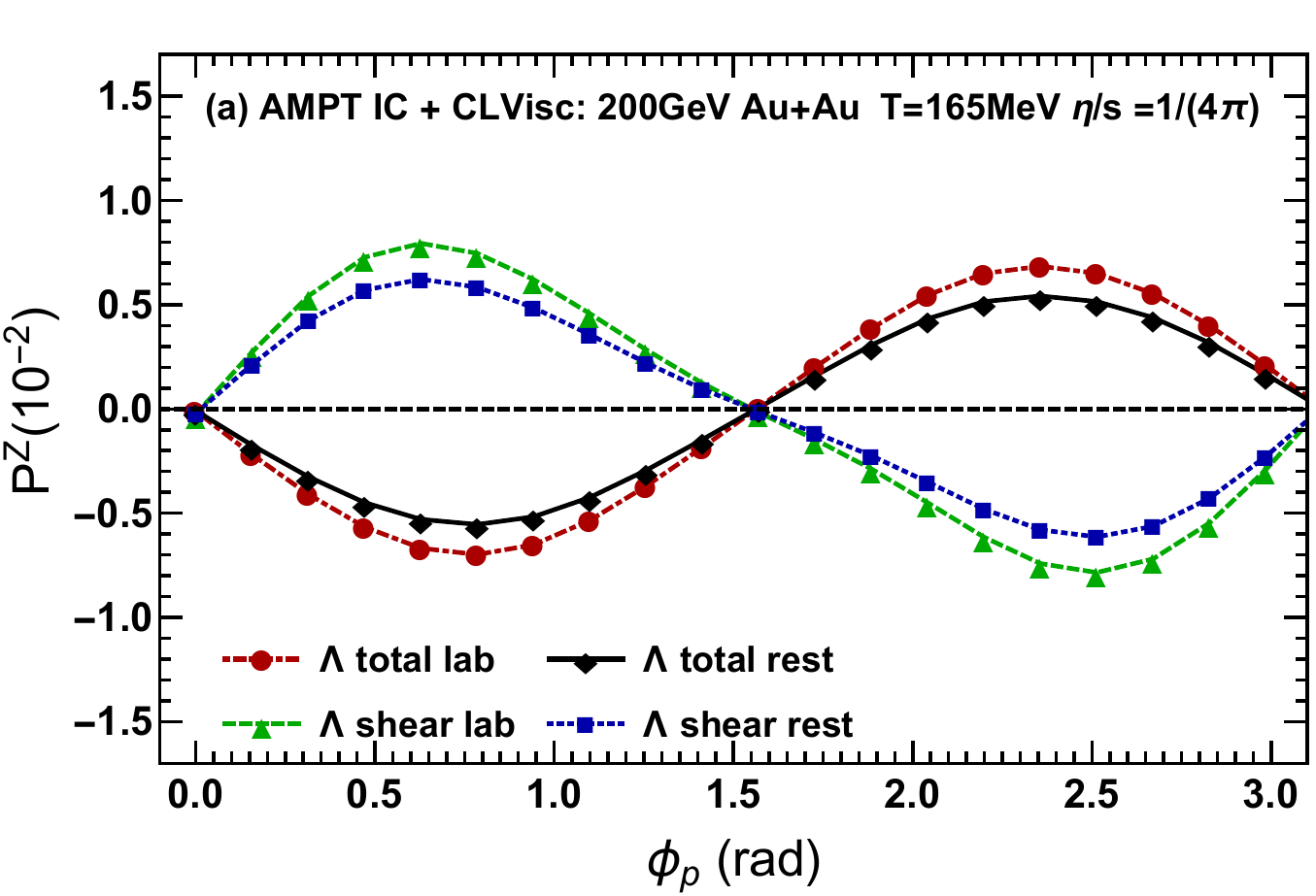}\includegraphics[scale=0.35]{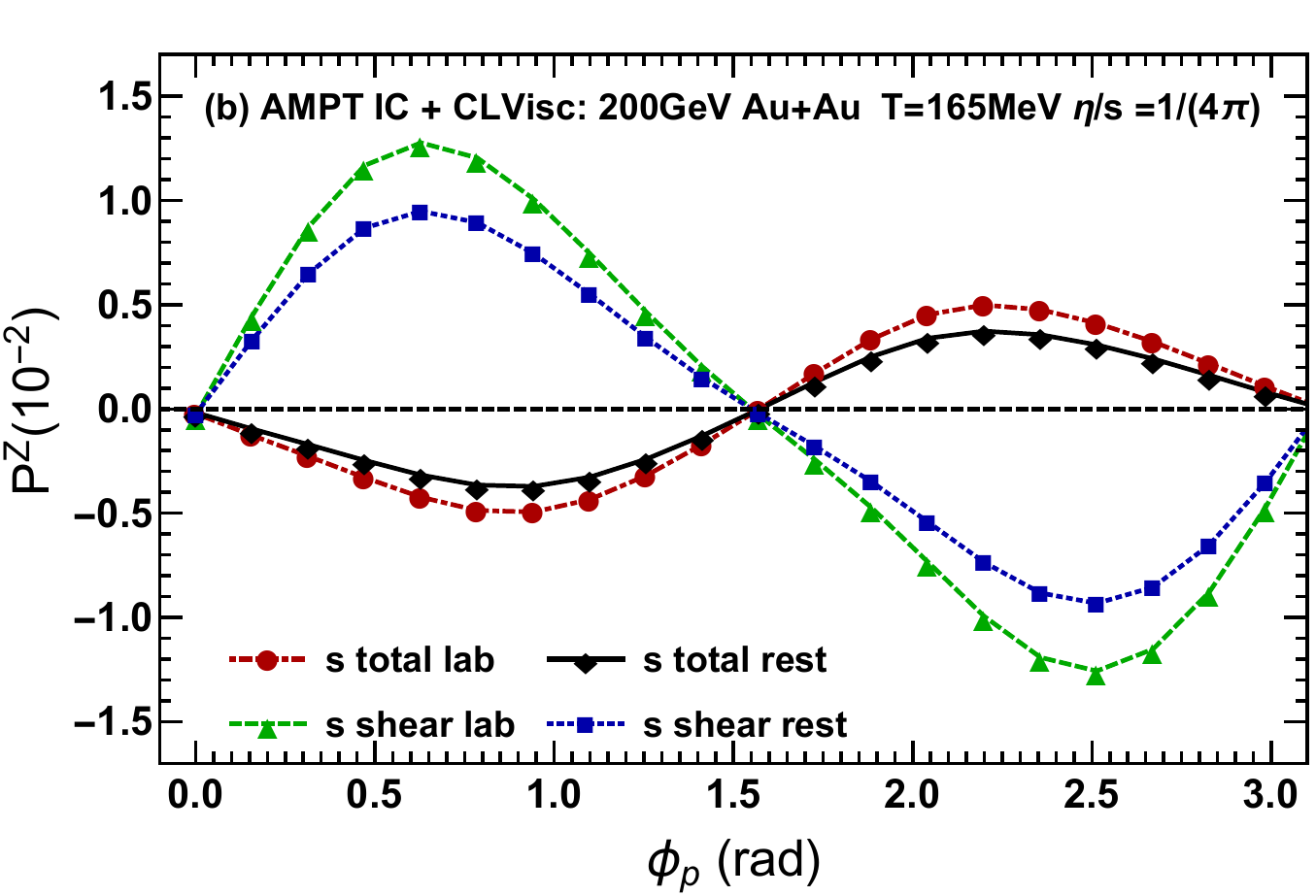}

\includegraphics[scale=0.35]{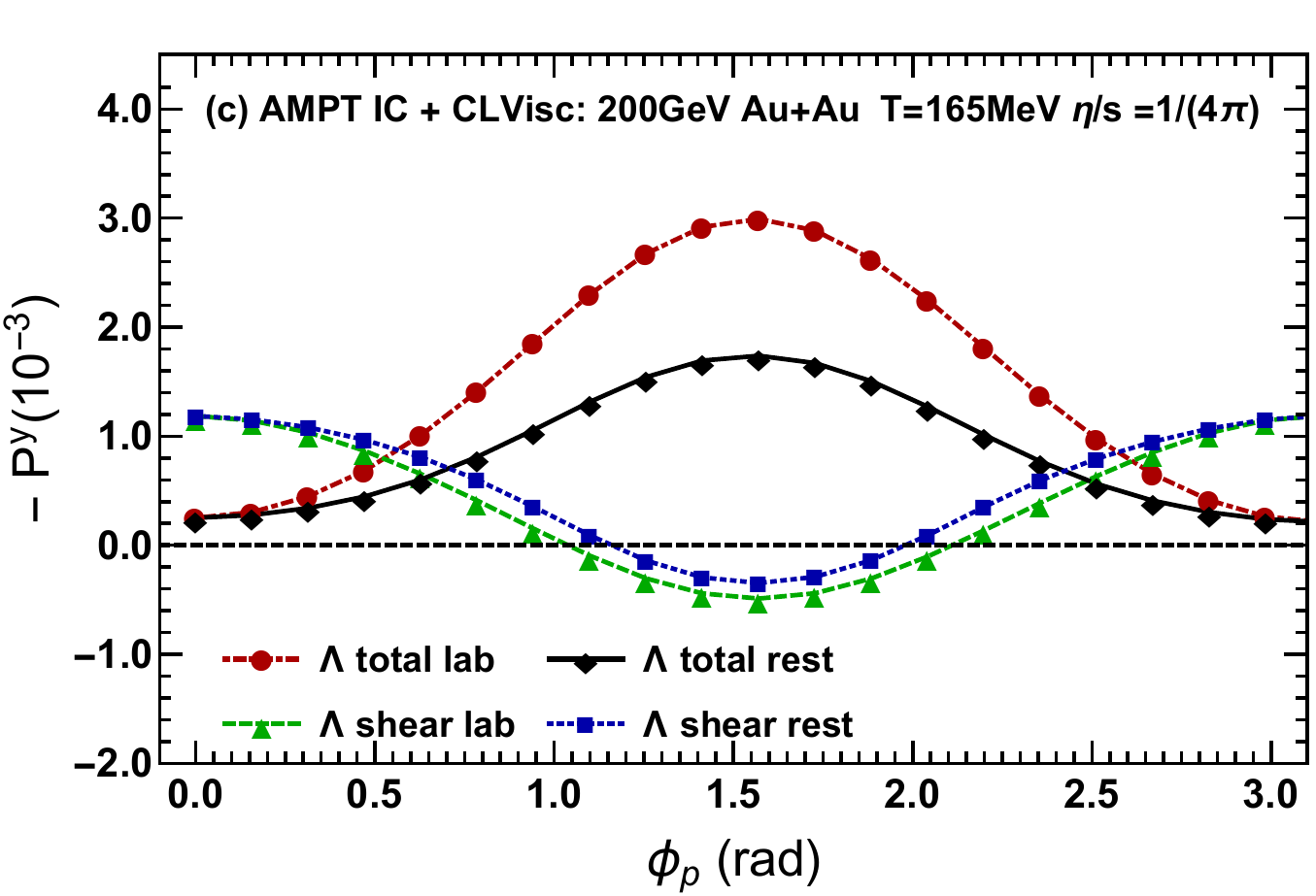}\includegraphics[scale=0.35]{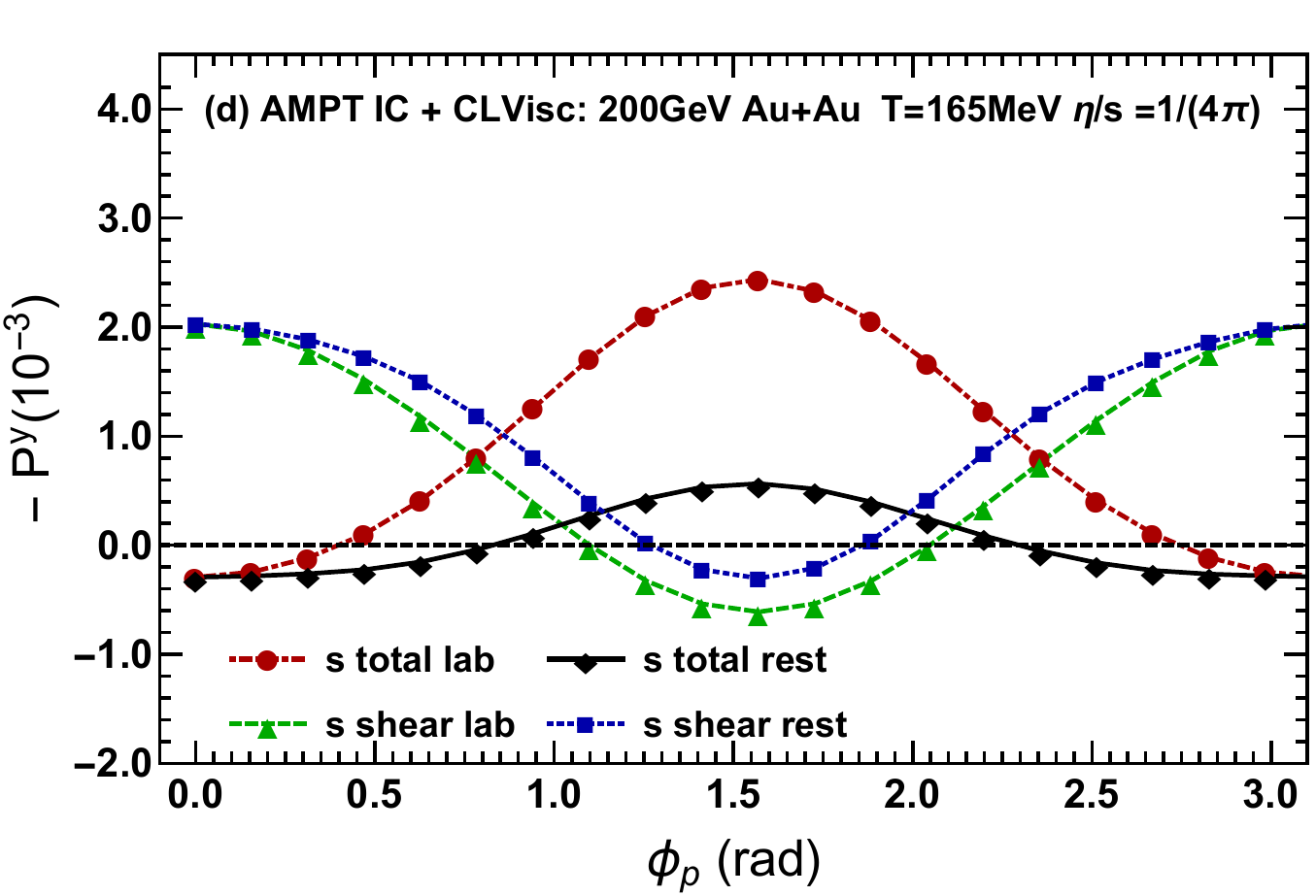}

\caption{The polarization $\mathcal{P}^{z}$ and $\mathcal{P}^{y}$ as a function
of $\phi_{p}$ for $\Lambda$ hyperons and $s$ quarks in the laboratory
frame and their own rest frames. We have chosen $\eta/s=1/(4\pi)$
and the freeze-out temperature is $165\,\textrm{MeV}$. Red and black
curves represent the total polarization in the laboratory and the
$\Lambda$ (or $s$ quark) rest frame, respectively. Green and blue
curves denote the shear induced polarization in the laboratory and
the $\Lambda$ (or the $s$ quark) rest frames, respectively. \label{fig:PzPy_rest_01} }
\end{figure}

We introduce the polarization along the beam direction as $\mathcal{P}^{z}(p)$
and that along the out-of-plane direction as $\mathcal{P}^{y}(p)$,
they are obtained by integrating over the medium rapidity range $[-1,+1]$,
\begin{eqnarray}
\mathcal{P}^{z}(p) & = & \int_{-1}^{+1}dY\mathcal{S}^{z}(p),\nonumber \\
\mathcal{P}^{y}(p) & = & \int_{-1}^{+1}dY\mathcal{S}^{y}(p),\label{eq:P_02}
\end{eqnarray}
where $\mathcal{S}^{\mu}$ is given by Eqs. (\ref{eq:S_all}). We
again use the subscripts, ``thermal'', ``shear'', ``accT'',
``chemical'', and ``EB'' for $\mathcal{P}^{\mu}(p)$ with $\mu=y,z$
to specify the contributions to the polarization from the thermal
vorticity, shear viscous tensor, $Du_{\beta}-\frac{1}{T}\partial_{\beta}T$,
gradient of $\mu/T$, and electromagnetic fields, respectively.

Since the electromagnetic fields decay rapidly \citep{Bloczynski:2012en,Deng:2012pc,Roy:2015coa},
we can neglect the contributions from electromagnetic fields $\mathcal{S}_{\textrm{EB}}^{\mu}$
to the polarization vector $\mathcal{P}_{\textrm{EB}}^{\mu}$. Due
to the limitation of the EoS \emph{s95p-pce}, we cannot get sufficient
information on the chemical potential and its gradient. Therefore,
in the current study, we only consider the polarization induced by
the thermal vorticity $\mathcal{S}_{\textrm{thermal}}^{i}$, shear
viscous tensor $\mathcal{S}_{\textrm{shear}}^{i}$, and $\mathcal{S}_{\textrm{accT}}^{i}$
and evaluate 
\begin{equation}
\mathcal{P}_{\textrm{total}}^{\mu}=\mathcal{P}_{\textrm{thermal}}^{\mu}+\mathcal{P}_{\textrm{shear}}^{\mu}+\mathcal{P}_{\textrm{accT}}^{\mu}.
\end{equation}
The possible contribution from $\mathcal{S}_{\textrm{chemical}}^{\mu}$
is also briefly discussed later.

\begin{figure}
\includegraphics[scale=0.4]{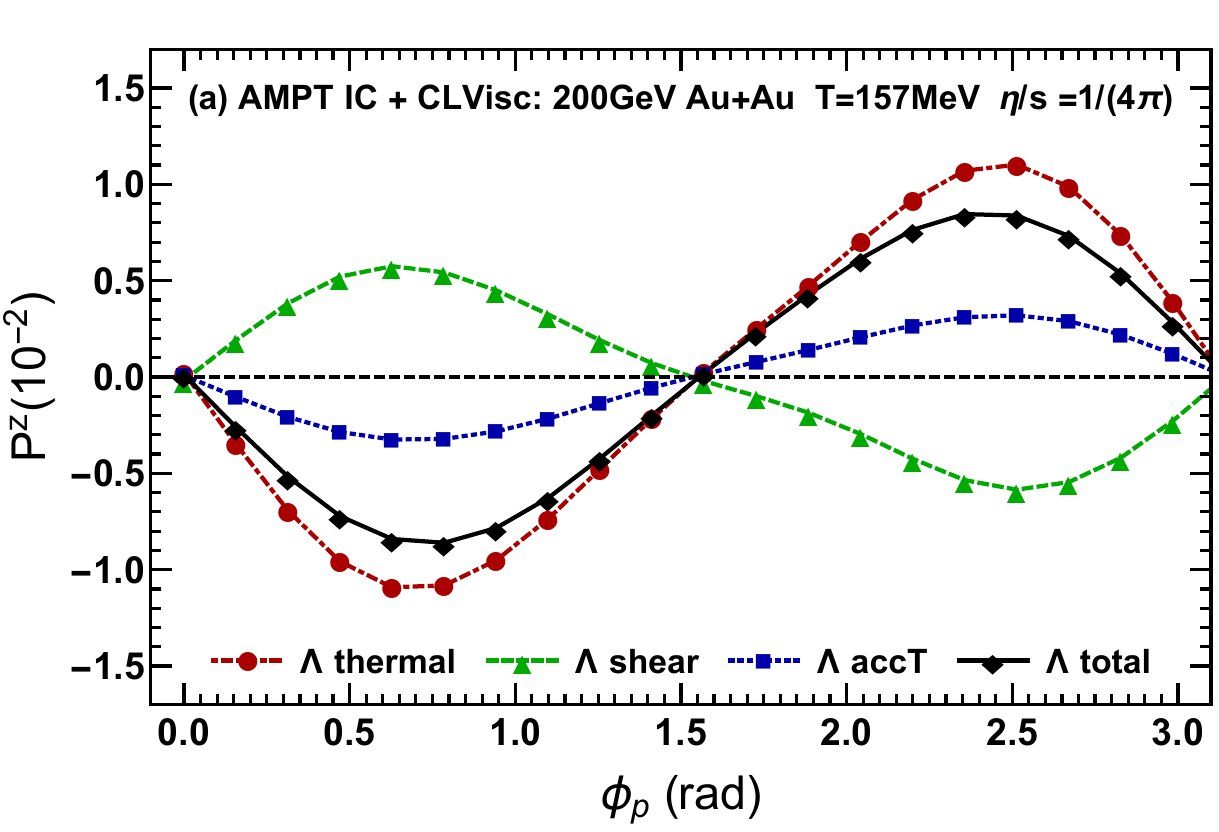}\includegraphics[scale=0.4]{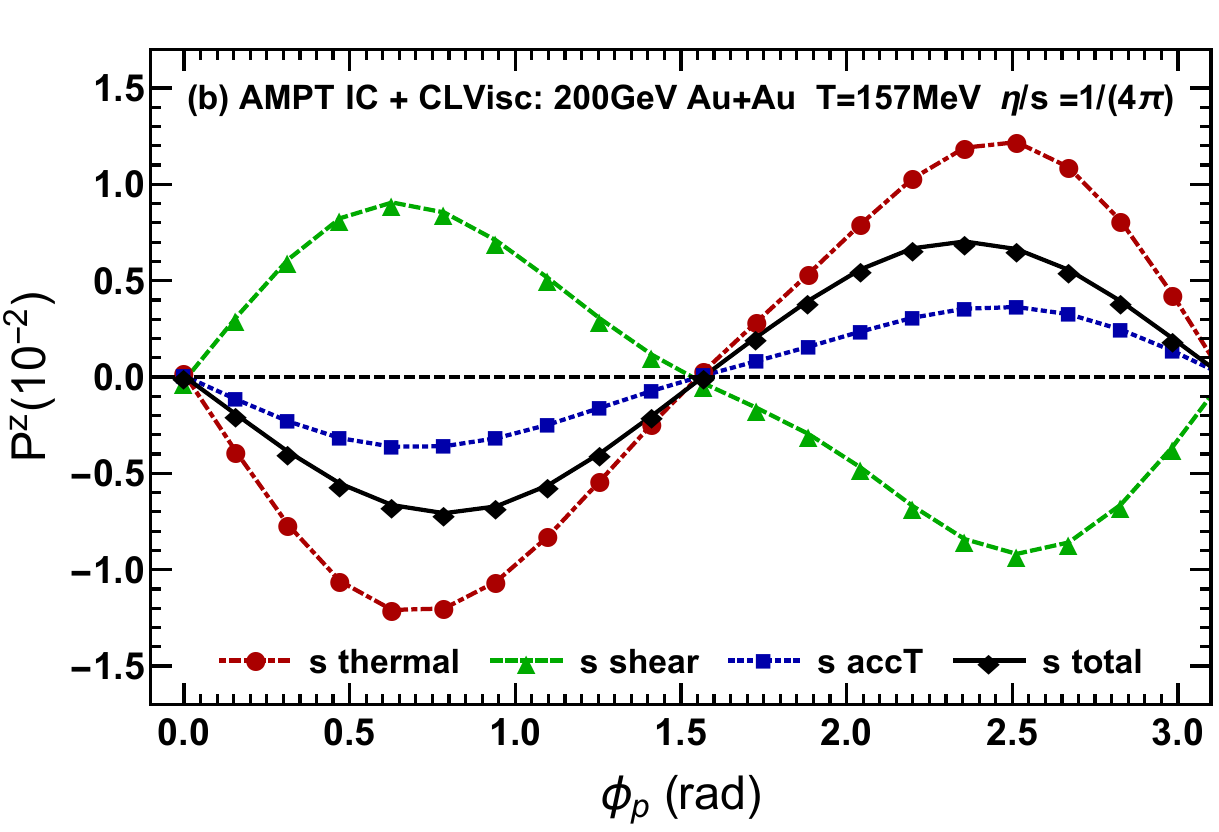}

\includegraphics[scale=0.4]{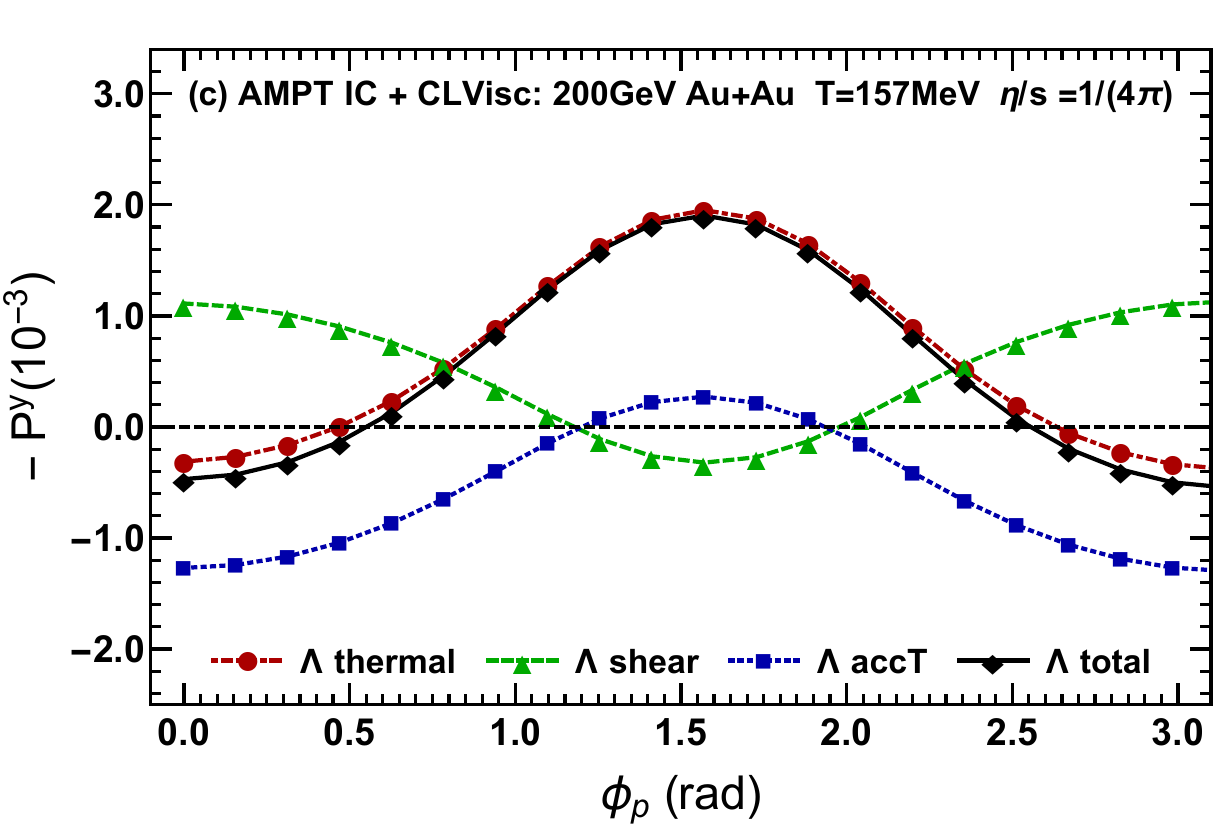}\includegraphics[scale=0.4]{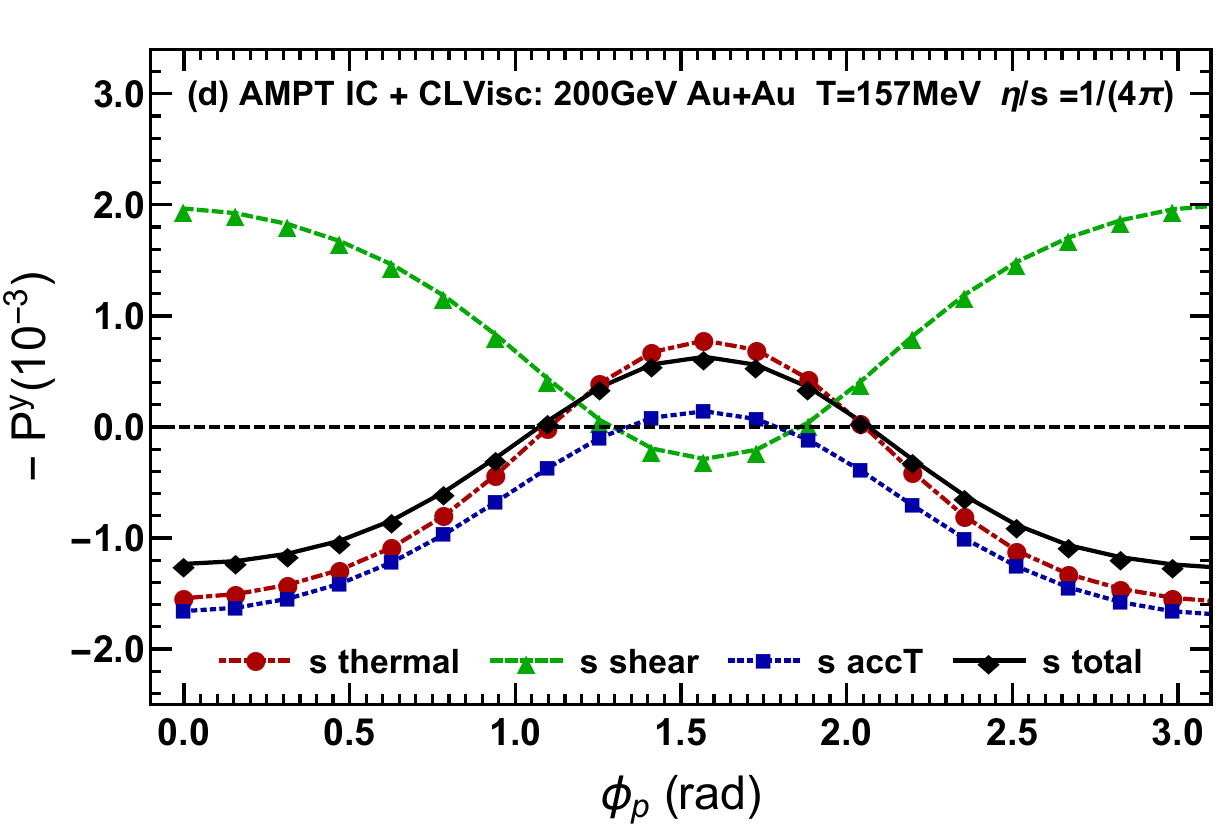}

\caption{
The same setup and color assignments as in Fig.~\ref{fig:PzPy_01}
except the freeze-out temperature has been changed to $157\,\textrm{MeV}$.
\label{fig:PzPy_T157} }
\end{figure}

\begin{figure}
\includegraphics[scale=0.4]{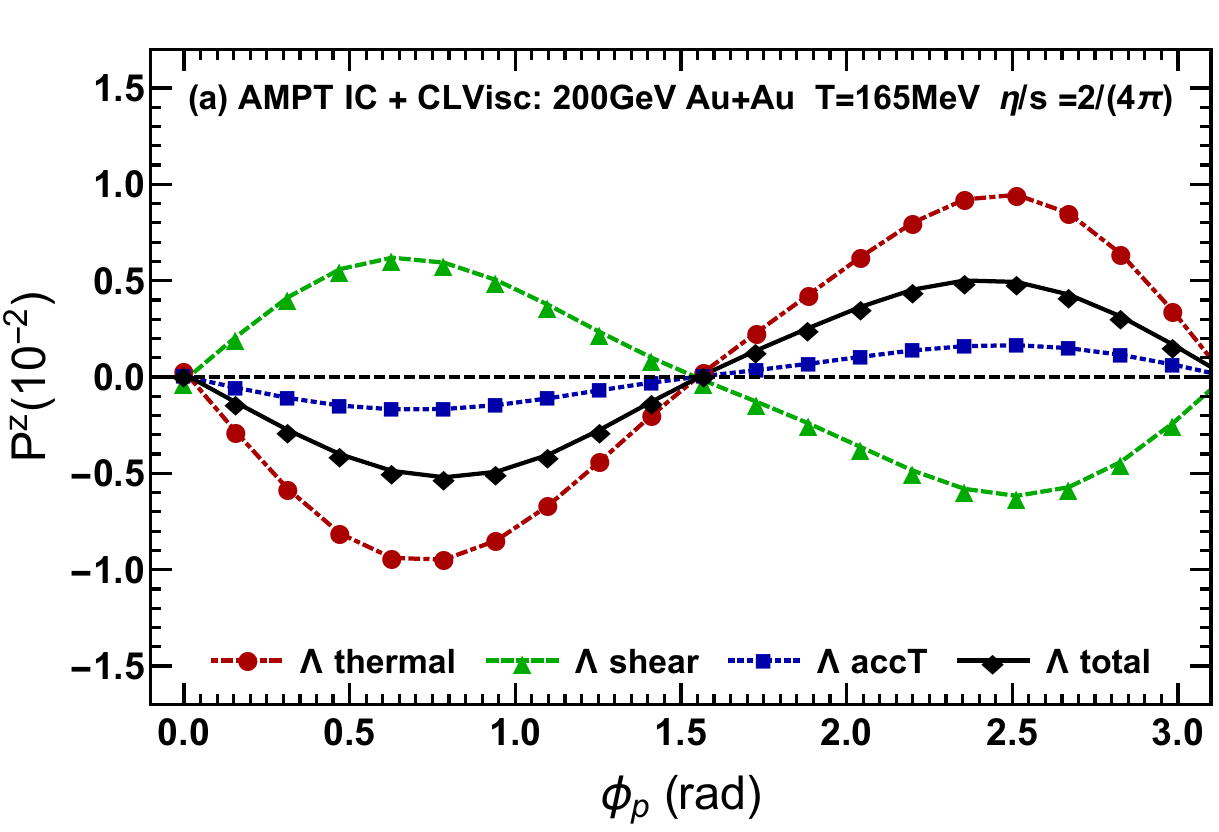}\includegraphics[scale=0.4]{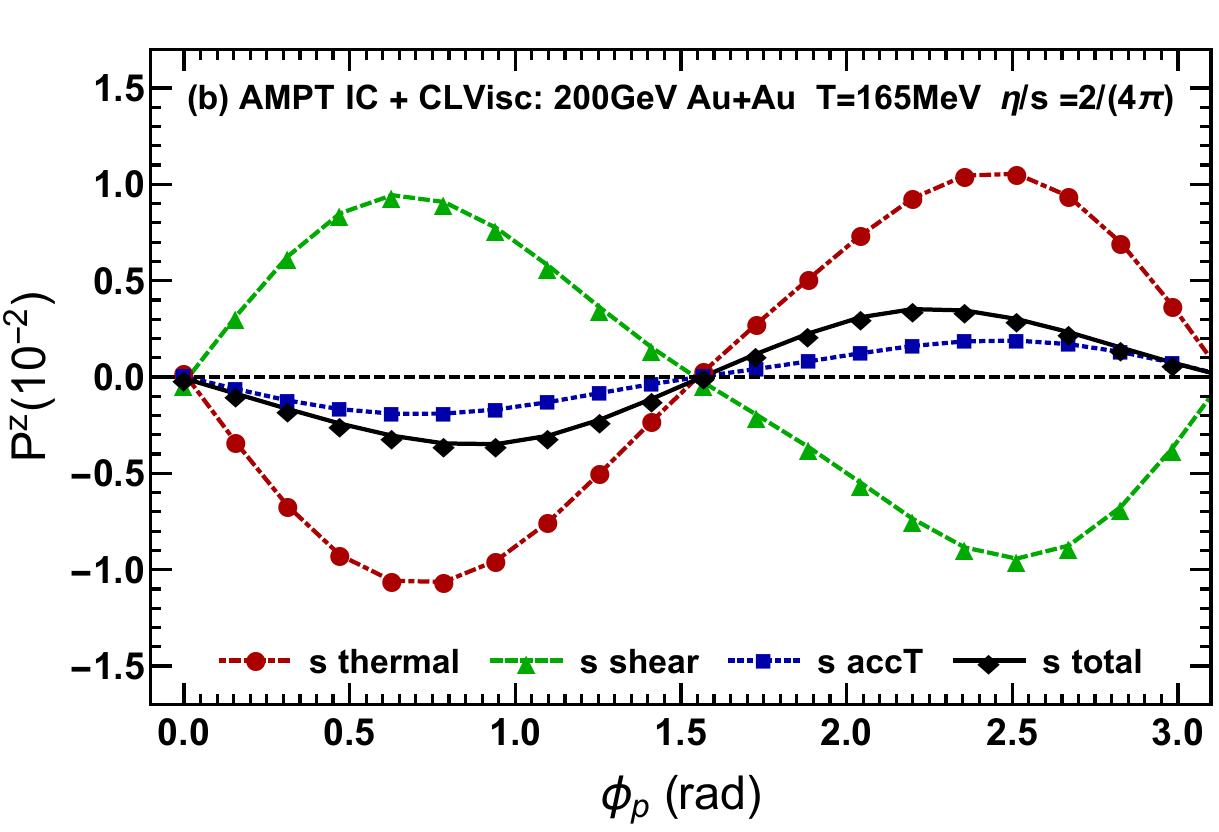}

\includegraphics[scale=0.4]{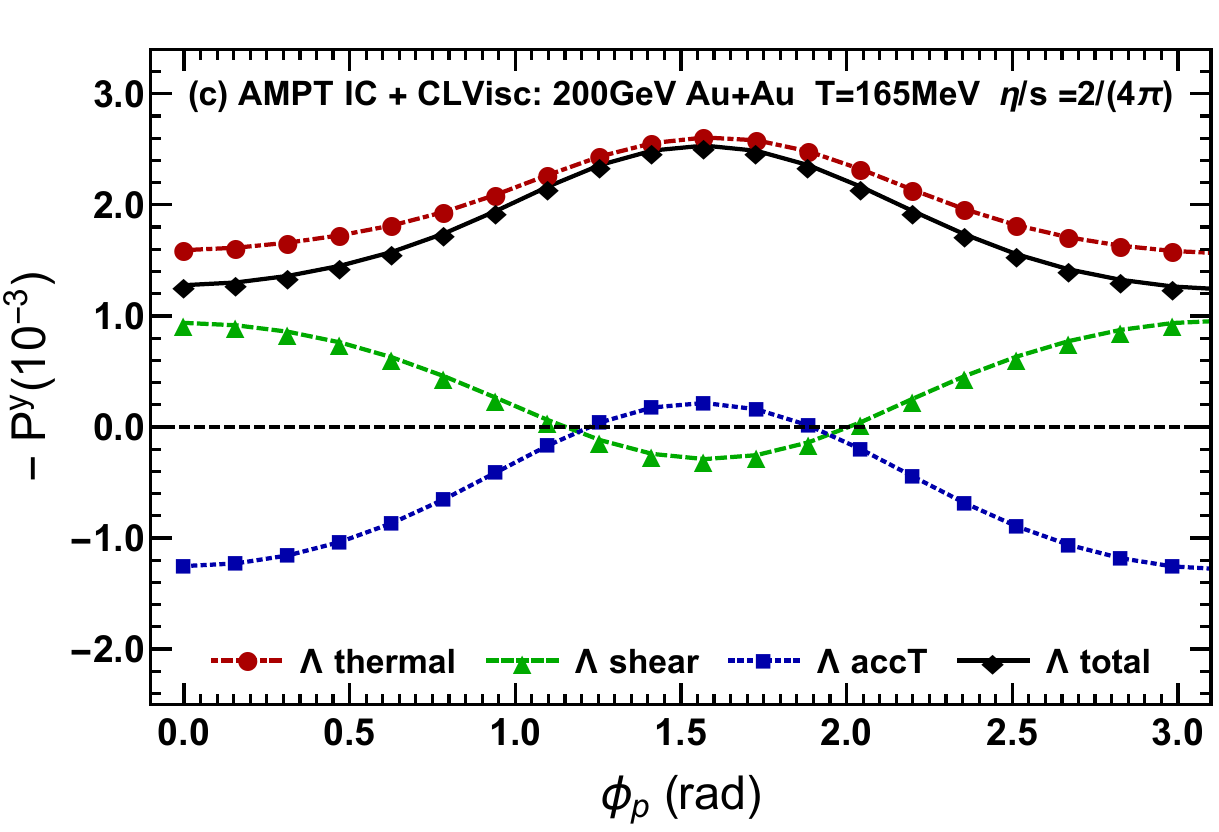}\includegraphics[scale=0.4]{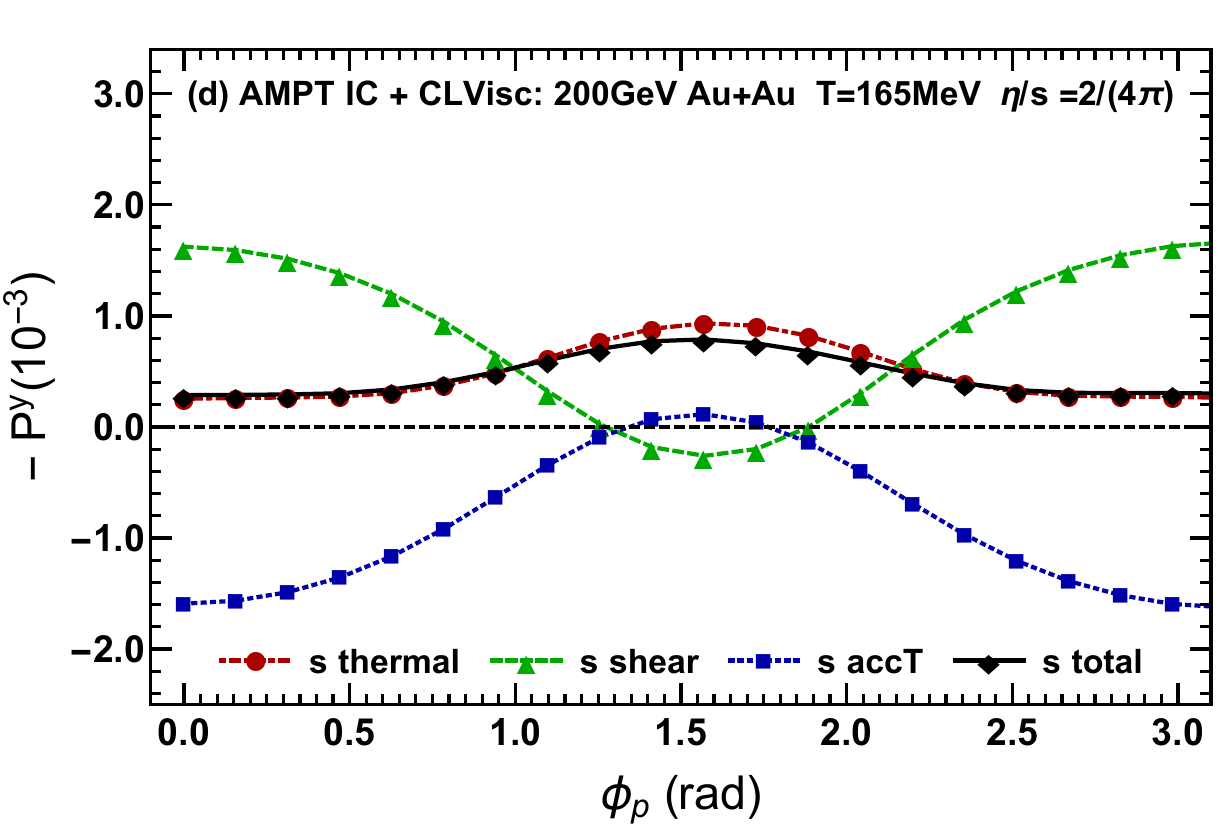}

\caption{
The same setup and color assignments as in Fig.~\ref{fig:PzPy_01}
except the shear viscosity-to-entropy density ratio has been changed
to $\eta/s=2/(4\pi)$. \label{fig:PzPy_eta016}}
\end{figure}

\begin{figure}[t]
\includegraphics[scale=0.35]{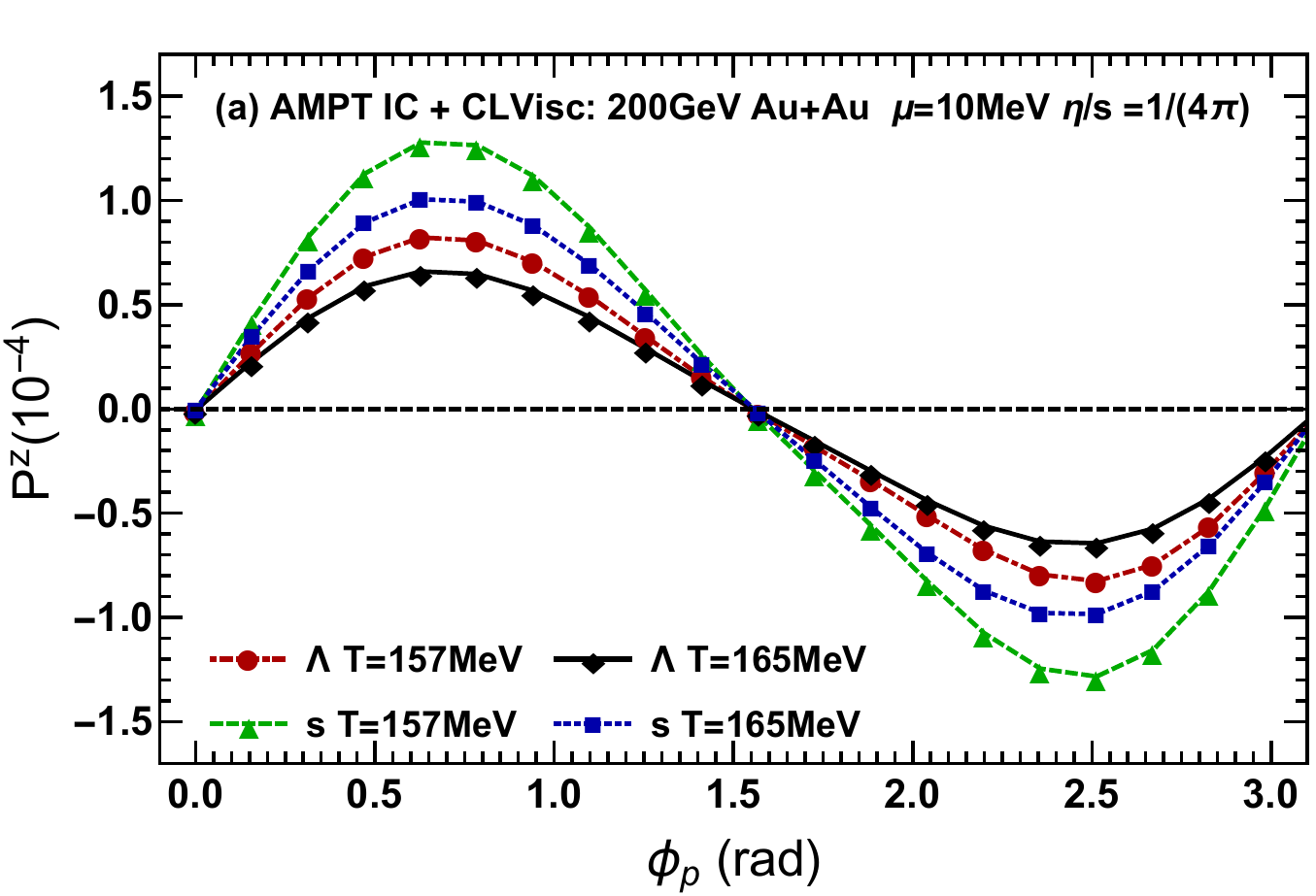}\includegraphics[scale=0.35]{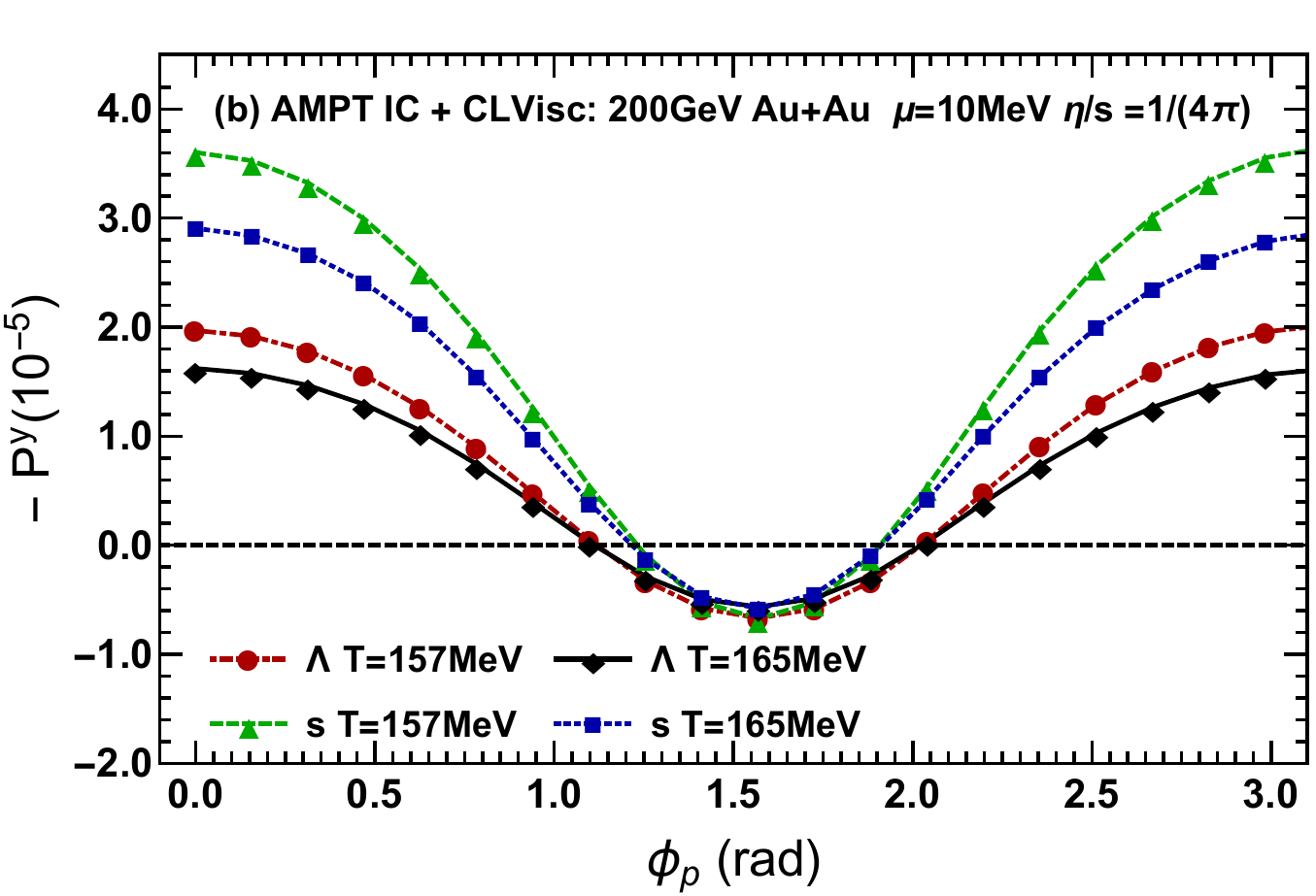}

\caption{The polarization $\mathcal{P}_{\textrm{chemcial}}^{z}$ and $\mathcal{P}_{\textrm{chemcial}}^{y}$
as a function of $\phi_{p}$ for the $\Lambda$ and $s$ equilibrium
scenarios. We have chosen $\eta/s=1/(4\pi)$ and the freeze-out temperature
is $165$ or $157$ MeV. See the color assignments for the results
in the plots. \label{fig:PzPy_eta016-1} }
\end{figure}

We consider two scenarios in this work. In the first scenario, since
the $\Lambda$ hyperons are produced at the chemical freeze-out, we
assume that one can still utilize the macroscopic variables from hydrodynamics
to describe the thermodynamical states of $\Lambda$ hyperons, e.g.
the temperature and its gradient, i.e. we assume that $\Lambda$ hyperons
are almost at local equilibrium. We call this the $\Lambda$ \emph{equilibrium
scenario} for short.

In the second scenario, as proposed in Ref. \citep{Fu:2021pok}, since
the spin of the $s$ quark dominates the total spin of the $\Lambda$
hyperons in the parton model, one can compute the polarization of
$s$ quarks and assume that the spin polarization of the\emph{ s }quark
is smoothly passed to the polarization of $\Lambda$. We call this
the $s$ \emph{equilibrium scenario}. Since $s$ quarks are much lighter
than $\Lambda$ hyperons, polarization induced by the shear viscous
tensor will be greatly enhanced in the $s$ equilibrium scenario as
opposed to the $\Lambda$ equilibrium scenario, where $\mathcal{S}_{\textrm{shear}}^{\mu}$
is suppressed by $(u\cdot p)\sim m_{\Lambda}$ in the denominator
shown in Eq. (\ref{eq:S_all}). 

Note that the polarization vectors in Eq. (\ref{eq:P_02}) are shown
in the laboratory frame. In order to compare the results with the
experimental data, we eventually transform them to the rest frame
of the $\Lambda$ hyperon in the $\Lambda$ equilibrium scenario or
of the $s$ quark in the $\Lambda$ hyperon in the $s$ equilibrium
scenario. In principle, we also need to consider the evolution of
$\Lambda$ hyperons before kinetic freeze-out, whereas we neglect
the evolution of $\Lambda$ hyperons after the chemical freeze-out
for simplicity.

We would like to emphasize that we always choose the factor $m$ in
the denominator on the right-hand side of Eq. (\ref{eq:S_01}) as
$m_{\Lambda}$ in both the $\Lambda$ and the $s$ equilibrium scenarios.
The mass factor in Ref. \citep{Fu:2021pok} is instead chosen as $m_{s}$
in the $s$ equilibrium scenario, which will enhance  the overall
magnitude of polarization in the $s$ equilibrium scenario. 
The reason that we still choose overall factor $m$ in in the denominator
on the right-hand side of Eq. (\ref{eq:S_01}) as $m_{\Lambda}$ in
the $s$ equilibrium scenario is as follows. Although it is assumed
that polarization of the $s$ quark dominates over other contributions
in a $\Lambda$ hyperon in the $s$ equilibrium scenario, $\mathcal{S}^{\mu}({\bf p})$
still corresponds to the polarization of the $\Lambda$ hyperon instead
of the $s$ quark. Nevertheless, this choice only changes the magnitude
of polarizations and does not change the qualitative results or main
conclusion. For numerical simulations, we choose $m_{\Lambda}=1.116\,\textrm{GeV}$
for the mass of $\Lambda$ hyperons and $m_{s}=0.3\,\textrm{GeV}$
for the mass of constituent $s$ quarks.

\subsection{Numerical Results and Discussions}

Here we briefly summarize the results presented in each figure. In
Fig. \ref{fig:PzPy_01}, we show the local spin polarization coming
from different sources in the local rest frame. In Fig.~\ref{fig:PzPy_rest_01},
we compare the results in the laboratory frame to those in Fig.~\ref{fig:PzPy_01}.
In Fig.~\ref{fig:PzPy_T157}, we present the results with a reduced
freeze-out temperature. In Fig.~\ref{fig:PzPy_eta016}, the results
with a larger $\eta/s$ but the same freeze-out temperature in Fig.~\ref{fig:PzPy_01}
are presented. Finally, we show the spin polarization led by a small
and constant chemical potential from $\mathcal{P}_{{\rm chemical}}^{\mu}$
for a heuristic discussion. The detailed results and discussion are
presented below.

In Fig. \ref{fig:PzPy_01}, we plot the local spin polarization in
both the $\Lambda$ and the $s$ equilibrium scenarios with $\eta/s=1/(4\pi)$
and the freeze-out temperature $T=165$ MeV. We observe that $\mathcal{\mathcal{S}_{\textrm{shear}}^{\mu}}$
always leads to the ``same'' sign contribution, qualitatively consistent
with experimental data, to the local spin polarization both in the
$z$ and $y$ directions and in the two scenarios. This observation
is in accordance with Ref. \citep{Fu:2021pok}. Similarly to the thermal
vorticity, $\mathcal{S}_{\textrm{accT}}^{\mu}$ induced by a non-vanishing
$Du_{\mu}-(\Delta_{\mu\nu}\partial^{\nu}T)/T$ leads to the ``opposite''
sign contribution to the local polarization. Along the beam direction,
the magnitude of polarization $\mathcal{P}_{\textrm{accT}}^{z}$ is
much smaller than the magnitudes of $\mathcal{P}_{\textrm{thermal }}^{z}$
and $\mathcal{P}_{\textrm{shear }}^{z}$ in the two scenarios. Along
the out-of-plane direction, the $\mathcal{P}_{\textrm{accT}}^{y}$
is almost canceled by $\mathcal{P}_{\textrm{shear}}^{y}$ in the $\Lambda$
equilibrium scenario, while $|\mathcal{P}_{\textrm{accT}}^{y}|<|\mathcal{P}_{\textrm{shear}}^{y}|$
in the $s$ equilibrium scenario. As a consequence of the competition
between $\mathcal{P}_{\textrm{shear}}^{\mu}$ and $\mathcal{P}_{\textrm{thermal}}^{\mu}+\mathcal{P}_{\textrm{accT}}^{\mu}$,
the $\mathcal{P}_{\textrm{total}}^{z,y}$ in both the $\Lambda$ and
the $s$ equilibrium scenarios disagree with the experimental data.

In contrast, the hydrodynamic simulations in Ref. \citep{Fu:2021pok}
find that the $\mathcal{P}_{\textrm{total}}^{\mu}$ in the $s$ equilibrium
scenarios agree with the experimental data qualitatively. As mentioned
previously, we have chosen a different EoS than the one in Ref. \citep{Fu:2021pok}.

One possible reason leading to the different results may derive from
the following fact. In the EoS \emph{sp95-pce}, the speed of sound
connecting the QGP phase and PCE hadronic phase is not smooth and
may lead to an un-physical vorticity structure near the freeze-out
surface. We note that the authors in Ref. \cite{Fu:2020oxj} have
studied the EoS dependence on the global polarization.

We have also checked the results by using another EoS \emph{lattice-wb2014}
\cite{Borsanyi:2013bia} and the total polarization has a similar
``sign'' as the experimental data qualitatively. But, $\mathcal{P}_{\textrm{total}}^{z,y}$
as a function of $\phi_{p}$ are almost flat and $|\mathcal{P}_{\textrm{total}}^{z}|$
is quantitatively close to 0 due to the high suppression led by the
factor $m_{\Lambda}$ in the denominator on the right-hand side of
Eq. (\ref{eq:S_01}). Therefore, our results show that the spin polarization
with local-equilibrium corrections from hydrodynamic simulations are
sensitive to the choice of EoS.

In Fig. \ref{fig:PzPy_rest_01}, we compare the polarization in the
laboratory frame and in the particles' rest frames. It is found that
the difference between the polarization in the two frames is slight
for $\mathcal{P}_{\textrm{total}}^{z}$. 
Although the overall magnitudes and peaks of $\mathcal{P}_{\textrm{total}}^{y}$
are reduced in the particles' rest frames, in general, the choices
of frames do not change the polarization qualitatively.

Next, we consider the dependence of the freeze-out temperature. We
set the freeze-out temperature $T=157$ MeV, keep $\eta/s=1/(4\pi)$,
and show the results in Fig. \ref{fig:PzPy_T157}. The magnitudes
of $\mathcal{P}_{\textrm{accT}}^{\mu}$ increase significantly when
the freeze-out temperature decreases \footnote{In fact, the magnitude of $\mathcal{P}_{\textrm{thermal}}^{\mu}$
also depends on the freeze-out temperature since $\mathcal{J}_{\textrm{thermal}}^{\mu}$
implicitly incorporates the term related to $Du^{\mu}$. But this
term has been previously included in the hydrodynamic simulations
for $\mathcal{P}_{\textrm{thermal}}^{\mu}$ at global equilibrium.
It is noteworthy that the full Wigner function at local equilibrium,
derived in Ref.~\cite{Hidaka:2017auj}, actually does not contain
a term associated with $Du^{\mu}$.}. Since now $|\mathcal{J}_{\textrm{accT}}^{\mu}|\sim\mathcal{O}\big(T^{-2}p\partial^{2}\big)$
is a crude estimation for the magnitude of $\mathcal{J}_{\textrm{accT}}^{\mu}$
coming from the dissipative correction, it may be expected that $|\mathcal{P}_{\textrm{thermal}}^{\mu}|$
should increase with the reduced freeze-out temperature. 

On the other hand, we observe that the shear-induced polarization
$\mathcal{P}_{\textrm{shear}}^{\mu}$ is not sensitive to the freeze-out
temperature, while the magnitude of the thermal vorticity induced
polarization $\mathcal{P}_{\textrm{thermal}}^{\mu}$ is slightly enhanced.
The difference between $\mathcal{P}_{\textrm{total}}^{\mu}$ in the
two scenarios and experimental data increases due to the increase
in the magnitude of $\mathcal{P}_{\textrm{accT}}^{\mu}$. It turns
out that $\mathcal{P}_{\textrm{total}}^{\mu}$ is also sensitive to
the freeze-out temperature.

Third, we discuss the dependence of $\eta/s$. In Fig. \ref{fig:PzPy_eta016},
we set $\eta/s=2/(4\pi)$ and the freeze-out temperature $T=165$
MeV. In comparison with the case where $\eta/s=1/(4\pi)$ in Fig.
\ref{fig:PzPy_01}, we find that none of the contributions to $\mathcal{P}^{z}$
in the two scenarios have changed much. On the contrary, the magnitudes
of both $\mathcal{P}_{\textrm{thermal}}^{y}$ and $\mathcal{P}_{\textrm{accT}}^{y}$
in the two scenarios increase when $\eta/s$ grows. The magnitudes
of $\mathcal{P}_{\textrm{shear}}^{y}$ in the two scenarios become
smaller than those in Fig. \ref{fig:PzPy_01}. Eventually, the $\mathcal{P}_{\textrm{total}}^{y}$
values in the two scenarios are still different from the observations
in experiments. It is found that the local polarization, at least
for $\mathcal{P}_{\textrm{total}}^{y}$, also depends on $\eta/s$.

At last, we compute the possible contribution from the gradient of
the chemical potential over the temperature led by $\mathcal{S_{\textrm{chemical}}^{\mu}}$
in Eq. (\ref{eq:S_all}). For simplicity, we assume that the chemical
potential is constant near the freeze-out hypersurface and $\nabla\mu/T\simeq\mu\nabla(1/T)$.
In Fig. \ref{fig:PzPy_eta016-1}, we choose the quark chemical potential
$\mu=10$ MeV. Similarly to the shear viscous tensor, $\mathcal{S_{\textrm{chemical}}^{\mu}}$
leads to the ``same'' sign contribution, qualitatively consistent
with the experimental data, to the local spin polarization both in
the $z$ and $y$ directions and in the two scenarios. However, $\mathcal{P}_{\textrm{chemical}}^{\mu}\propto\mu$
is greatly suppressed by other contributions to $\mathcal{P}_{\textrm{total}}^{\mu}$.
So far, they are almost negligible in the current study. However,
$\mathcal{P_{\textrm{chemical}}^{\mu}}$ might be important in low-energy
collisions. We leave this for future studies.

Before ending this section, we also comment on the choice of $m_{s}$.
So far, there are no first principle calculations to demonstrate how
large $m_{s}$ should be in these simulations. In general, one can
set $m_{s}$ as the current quark mass or constituent quark mass different
from $0.3$ GeV. As discussed in this section, we have shown that
$\mathcal{P}_{\textrm{shear}}^{\mu}$ competes with $\mathcal{P}_{\textrm{thermal}}^{\mu}$
and $\mathcal{P}_{\textrm{accT}}^{\mu}$. From Eq.~(\ref{eq:S_all}),
we find that the denominator $(u\cdot p)\sim p^{0}\sim m_{i}$ in
$\mathcal{S}_{\textrm{shear}}^{\mu}$ is sensitive to the mass $m_{i}$.
Since $m_{i}=m_{s}$ in the $s$ equilibrium scenario is much smaller
than $m_{i}=m_{\Lambda}$ in the $\Lambda$ equilibrium scenario,
$\mathcal{P}_{\textrm{shear}}^{\mu}$ is therefore significantly enhanced
in the $s$ equilibrium scenario. Moreover, if we increase $m_{s}$,
$\mathcal{P}_{\textrm{shear}}^{\mu}$ obviously decreases. A more
rigorous study of the spin fragmentation from \emph{s }quarks to $\Lambda$
hyperons in the hadronization process is required to resolve this
uncertainty in the future. 

As a concluding remark, we have shown that $\mathcal{P}_{\textrm{total}}^{\mu}$
is generally sensitive to the EoS, freeze-out temperature $T$, and
ratio $\eta/s$. We find that both $\mathcal{S}_{\textrm{shear}}^{\mu}$
and $\mathcal{S}_{\textrm{accT}}^{\mu}$ play important roles in local
spin polarization. The behavior of $\mathcal{P}_{\textrm{total}}^{\mu}$
greatly depends on the predominant term in the competition between
$\mathcal{P}_{\textrm{shear}}^{\mu}$ and $\mathcal{P}_{\textrm{thermal}}^{\mu}+\mathcal{P}_{\textrm{accT}}^{\mu}$.
Although $\mathcal{S}_{\textrm{shear}}^{\mu}$ may lead to the ``same''
sign of the polarization, we still cannot get a similar azimuthal
angle dependence of the local spin polarization as the experimental
data with the EoS we have chosen.

\section{Conclusion \label{sec:Conclusion}}

In this work, we first review the spin polarization (pseudo-) vector
$\mathcal{S}^{\mu}({\bf p})$ derived from Wigner functions in Ref.~\citep{Hidaka:2017auj}.
We decompose $\mathcal{S}^{\mu}({\bf p})$ into $\mathcal{S}_{\textrm{thermal}}^{\mu},\mathcal{S}_{\textrm{shear}}^{\mu},\mathcal{S}_{\textrm{accT}}^{\mu},\mathcal{S}_{\textrm{chemical}}^{\mu}$
and $\mathcal{S}_{\textrm{EB}}^{\mu}$, which are led by the thermal
vorticity, shear viscous tensor, fluid acceleration minus the gradient
of temperature, gradient of the chemical potential over the temperature,
and electromagnetic fields, respectively. 
We then implement (3+1)-dimensional viscous hydrodynamic CLVisc with
AMPT initial conditions to obtain the numerical results for local
spin polarization with zero electromagnetic fields and a vanishing
chemical potential and focus on $\mathcal{P}_{\textrm{thermal}}^{\mu},\mathcal{P}_{\textrm{shear}}^{\mu}$,
and $\mathcal{P}_{\textrm{accT}}^{\mu}$ contributed by $\mathcal{S}_{\textrm{thermal}}^{\mu},\mathcal{S}_{\textrm{shear}}^{\mu}$,
and $\mathcal{S}_{\textrm{accT}}^{\mu}$. Inspired by Ref. \citep{Fu:2021pok},
we also consider $\Lambda$ and $s$ equilibrium scenarios. As opposed
to Ref.~\citep{Fu:2021pok}, we have chosen the EoS ``\emph{s95p-pce}''
as a different EoS and investigate the influence of $\eta/s$ upon
spin polarization. 

Our numerical results show that both $\mathcal{P}_{\textrm{shear}}^{\mu}$
and $\mathcal{P}_{\textrm{accT}}^{\mu}$ affect $\mathcal{P}_{\textrm{total}}^{\mu}$
in addition to $\mathcal{P}_{\textrm{thermal}}^{\mu}$, which has
been well studied in the literature. Although $\mathcal{P}_{\textrm{shear}}^{\mu}$
can lead to the ``same'' sign contribution, qualitatively consistent
with experimental data, to the local spin polarization, $\mathcal{P}_{\textrm{accT}}^{\mu}$
also plays a crucial role and could change the behavior of $\mathcal{P}_{\textrm{total}}^{\mu}$
especially with a lower freeze-out temperature. It turns out that
$\mathcal{P}_{\textrm{total}}^{\mu}$ is rather sensitive to EoS,
$\eta/s$ and freeze-out temperature $T$. With the EoS adopted in
our study, we do not observe a similar azimuthal angle dependence
of the local spin polarization as in the experimental data.

We conclude that although the shear-induced polarization alone may
result in sizable effects on the local spin polarization qualitatively
consistent with experimental observations, it may not always be the
dominant effect over other contributions at local equilibrium.

Although we have also checked the total local spin polarization by
using another EoS and gotten behavior similar to that in Ref.~\cite{Fu:2021pok},
the total local spin polarization in the\emph{ s }equilibrium scenario
could still be highly suppressed if one chooses $m_{\Lambda}$ instead
of $m_{s}$ in the denominator on the right-hand side of Eq.~(\ref{eq:S_01}).

We also note that Ref. \citep{Becattini:2021iol} added the contribution
from the shear viscous tensor to the local spin polarization and assumed
isothermal equilibrium, in which the gradient of the temperature is
assumed to be vanishing near the freeze-out hypersurface. It is also
tempting to study the dependence of the EoS near isothermal equilibrium
in the future.

Consequently, the so-called ``sign'' problem remains an open question.
Moreover, from the theoretical perspective, the Wigner function at
local equilibrium without the inclusion of off-equilibrium corrections
pertinent to interaction is not a self-consistent solution for a kinetic
equation. As indirectly supported by the influence of {$\mathcal{P}_{\textrm{accT}}^{\mu}$,
even under the near-equilibrium condition, the second-order gradient
terms, such as the non-equilibrium corrections qualitatively studied
in, e.g., Refs.~\cite{Hidaka:2017auj,Hidaka:2018ekt} with chiral
kinetic theory, may potentially give rise to sizable contributions
in numerical simulations. To comprehensively investigate the spin
polarization of fermions near local equilibrium and its direct connection
to the sign problem, it will still be imperative to conduct theoretical
and numerical studies of spin hydrodynamics and the QKT with collisions.
\begin{acknowledgments}
We are grateful to Yi Yin, Longgang Pang, Shuai Y. F. Liu, Baochi
Fu, Huichao Song and Francesco Becattini for helpful discussions.
S.P. was supported by the National Nature Science Foundation of China
(NSFC) under Grants No. 12075235 and No. 12135011. D.-L. Y. was supported
by the Ministry of Science and Technology, Taiwan, under Grant No.
MOST 110-2112-M-001-070-MY3. 
\end{acknowledgments}

 \bibliographystyle{h-physrev}
\bibliography{polarization}

{document} }
\end{document}